\documentclass[journal]{IEEEtran}
\usepackage{epsfig,amsmath,amssymb,color,subfigure,fmtcount,stfloats,color,bm}
\usepackage[noadjust]{cite}

\IEEEoverridecommandlockouts
\newcommand{\pr}{\text{P}}
\newcommand{\E}{\text{E}}
\newcommand{\ed}{\stackrel{\text{d}}{=}}
\newcommand{\deq}{\triangleq}

\newtheorem{definition}{Definition}
\newtheorem{theorem}{\textbf{Theorem}}
\newtheorem{lemma}{Lemma}
\newtheorem{remark}{Remark}

\makeatletter
\def\ps@headings{%
\def\@oddhead{\mbox{}\scriptsize\rightmark \hfil \thepage}%
\def\@evenhead{\scriptsize\thepage \hfil \leftmark\mbox{}}%
\def\@oddfoot{}%
\def\@evenfoot{}}
\makeatother
\begin{document}
\title{On the Critical Delays of Mobile Networks under L\'{e}vy Walks and L\'{e}vy Flights}

\author{Kyunghan Lee$^\dag$,~\IEEEmembership{Member,~IEEE},
Yoora Kim$^\ddag$,~\IEEEmembership{Member,~IEEE},
Song Chong$^\sharp$,~\IEEEmembership{Member,~IEEE}, \\
Injong Rhee$^\dag$,~\IEEEmembership{Member,~IEEE},
Yung Yi$^\sharp$,~\IEEEmembership{Member,~IEEE}, and
Ness. B. Shroff$^\ddag$,~\IEEEmembership{Fellow,~IEEE} \\
$^\dag$\{klee8, rhee\}@ncsu.edu, $^\ddag$\{kimy, shroff\}@ece.osu.edu, $^\sharp$\{songchong, yiyung\}@kaist.edu

\thanks{$\dag$: Department of Computer Science, North Carolina State University, USA. $\ddag$: Department of Electrical and Computer Engineering, The Ohio State University, USA $\sharp$: Department of Electrical Engineering, KAIST, South Korea. Part of the work done by Kyunghan Lee was performed while he was with KAIST. Yoora Kim performed her work in this paper when she was with KAIST. The order from the 3rd to the 5th authors is alphabetical.}
\thanks{A preliminary version of this paper was published in the Proceedings of the IEEE Conference on Computer Communications (INFOCOM), Shanghai, China, April 10-15, 2011.}}
\maketitle

\begin{abstract}
Delay-capacity tradeoffs for mobile networks have been analyzed through a number of research work. However, L\'{e}vy mobility known to closely capture human movement patterns has not been adopted in such work. Understanding the delay-capacity tradeoff for a network with L\'{e}vy mobility can provide important insights into understanding the performance of real mobile networks governed by human mobility. This paper analytically derives an important point in the delay-capacity tradeoff for L\'{e}vy mobility, known as the critical delay. The critical delay is the minimum delay required to achieve greater throughput than what conventional static networks can possibly achieve (i.e., $O(1/\sqrt{n})$ per node in a network with $n$ nodes). The L\'{e}vy mobility includes L\'{e}vy flight and L\'{e}vy walk whose step size distributions parametrized by $\alpha \in (0,2]$ are both heavy-tailed while their times taken for the same step size are different. Our proposed technique involves (i) analyzing the joint spatio-temporal probability density function of a time-varying location of a node for L\'{e}vy flight and (ii) characterizing an embedded Markov process in L\'{e}vy walk which is a semi-Markov process. The results indicate that in L\'{e}vy walk, there is a phase transition such that for $\alpha \in (0,1)$, the critical delay is always $\Theta (n^{\frac{1}{2}})$ and for $\alpha \in [1,2]$ it is $\Theta(n^{\frac{\alpha}{2}})$. In contrast, L\'{e}vy flight has the critical delay $\Theta(n^{\frac{\alpha}{2}})$ for $\alpha\in(0,2]$.
\end{abstract}

\section{Introduction}
Since the seminal work by Gupta and Kumar \cite{SL1Gupta} on the capacity of wireless networks, delay and throughput tradeoffs for wireless networks have been extensively studied for various mathematical techniques, scheduling algorithms, channel models, mobility models and physical layer techniques. The work by Grossglauser and Tse~\cite{grossglauser:mobility} showed that the per-node throughput remains constant ($\Theta(1)$) when node mobility is used for communication. This result is surprising because Gupta and Kumar \cite{SL1Gupta} had previously shown that the per-node throughput ($O(1/\sqrt{n})$) in wireless networks with no mobility diminishes as the number of nodes $n$ increases. This throughput gain is achieved at the cost of larger delays.

The amount of delay that a network needs to sacrifice to guarantee a given throughput has been studied under various mobility models~\cite{SL17Toumpis, SL14Neely, SL21Lin}. In particular, Sharma~\emph{et al.}~\cite{SL_Sharma07} studied the minimum delays required to achieve more per-node throughput than $\Theta(1/\sqrt{n})$\footnote{As \cite{SL1Gupta} showed, $\Theta(1/\sqrt{n})$ is the maximum throughput that wireless networks relying on naive multi-hop transmissions can achieve without the help of node mobility.}  under various mobility models including i.i.d., random waypoint, random direction and Brownian motion. This minimum delay is called {\it critical delay}. However, although the work provides a nice framework for studying delay-capacity scaling for wireless networks under a family of random walk models, the practical values of these mobility models are limited. While these models are simple enough for mathematical tractability, they do not reflect realistic mobility patterns commonly exhibited in real mobile networks.

\begin{figure}[t!]
  \centering
  \subfigure[Brownian motion]{\epsfig{figure=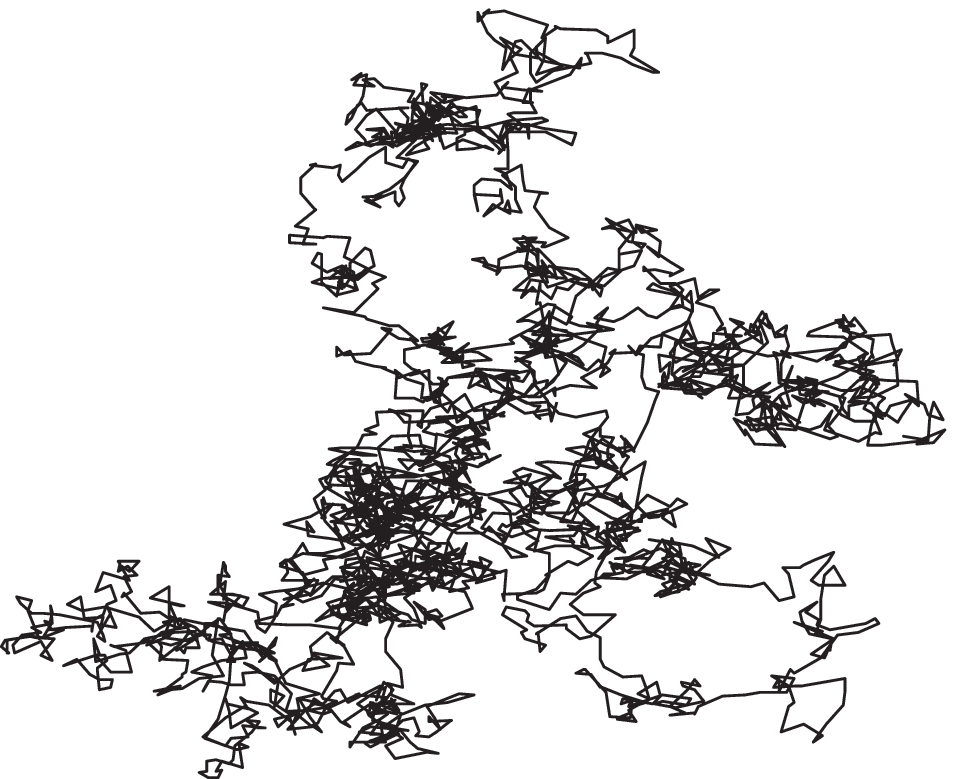,width=0.15\textwidth,height=0.15\textwidth}}
  \subfigure[L\'{e}vy
  mobility]{\epsfig{figure=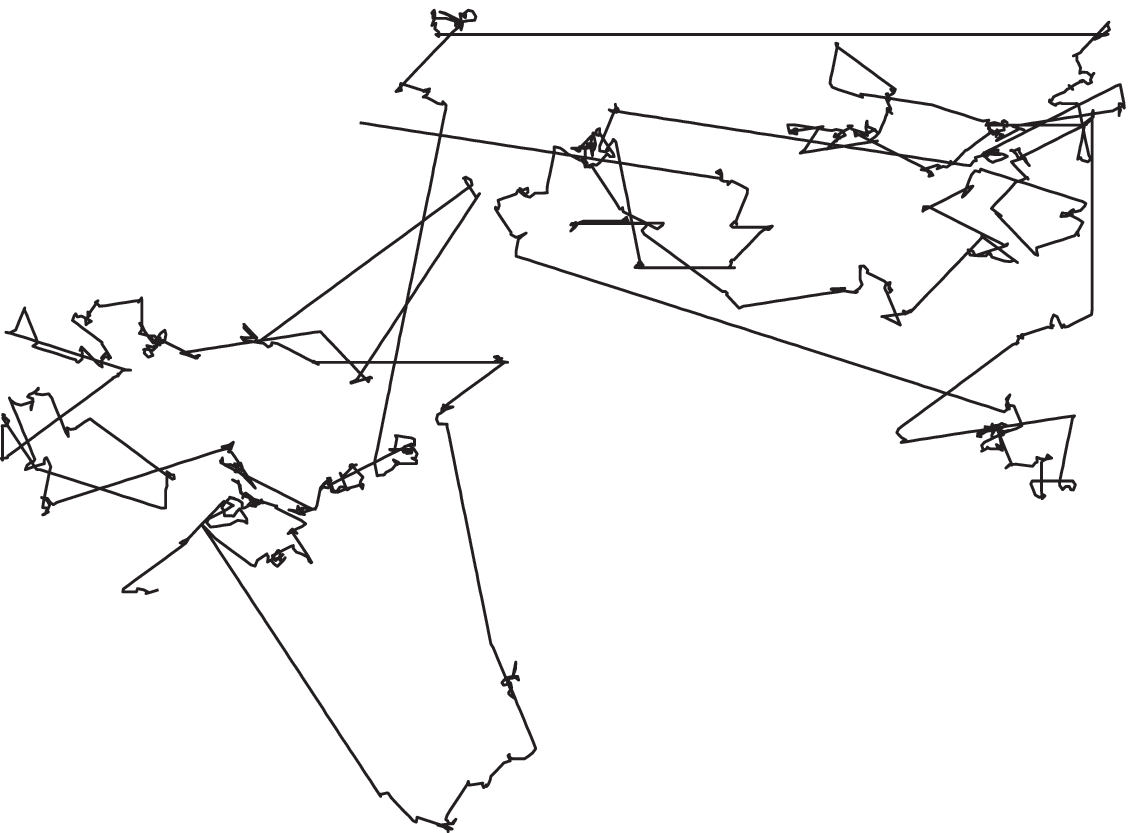,width=0.15\textwidth,height=0.15\textwidth}}
  \subfigure[Random waypoint]{\epsfig{figure=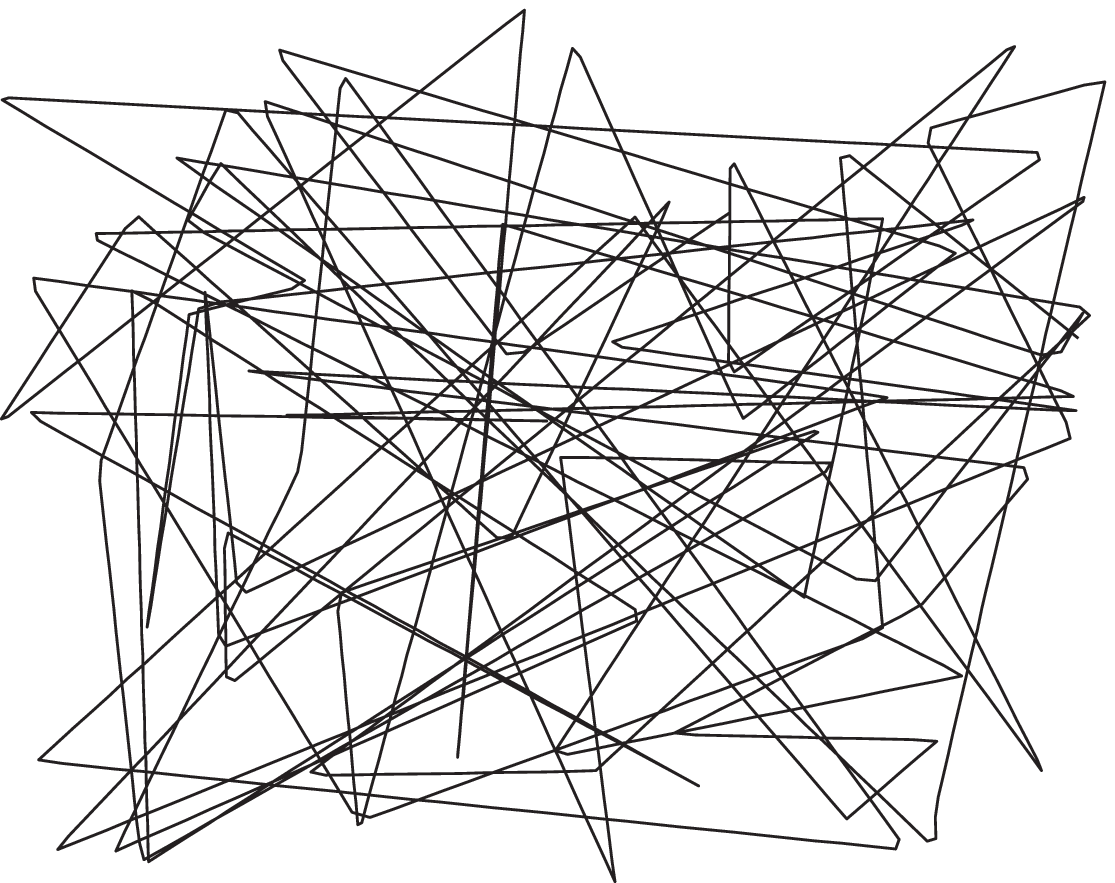,width=0.15\textwidth,height=0.15\textwidth}}
  \caption{Sample trajectories of (a) Brownian motion, (b) L\'{e}vy mobility and (c) random waypoint.}
  \label{fig:sampleTr}
\end{figure}

Humans are a major factor in mobile networks as most mobile nodes or devices (smartphones and cars) are carried or driven by humans. Recent studies  \cite{rhee:levymobility,lee:slaw09,gonalez:understanding} on human mobility show that step size distributions\footnote{Step size is often referred to as flight length in some literatures.} are heavy-tailed where {\em a step} is defined to be the straight line trip of a moving object (e.g., particles or humans) from one location to another without a directional change or pause. These mobility patterns are accurately modeled by L\'{e}vy process \cite{shlesinger:levyDynamics}.

L\'{e}vy mobility is a random walk mobility whose step size distribution is parametrized by $\alpha \in (0,2]$ and is heavy-tailed except in the extreme case of $\alpha =2$. For $\alpha \in (0,2)$, the distribution is well approximated by a power-law distribution $1/z^{1+\alpha}$ where $z$ is a step size. For $\alpha = 2$, the step size conforms to Gaussian distribution.\footnote{L\'{e}vy mobility becomes Brownian motion in the extreme case of $\alpha = 2$.} Intuitively, such a random walk contains many short steps and a small yet significant number of exceptionally long steps. With different values of~$\alpha$, the movement patterns of L\'{e}vy mobility models are widely different. Smaller $\alpha$ induces a larger number of long steps. This type of mobility patterns is significantly different from Brownian motion and random waypoint as illustrated in Fig.~\ref{fig:sampleTr}. In the literature, there are two types of L\'{e}vy mobility models for classification: {\em L\'{e}vy flight} and {\em L\'{e}vy walk}. In L\'{e}vy flight, every step takes a {\em constant time} irrespective of its step size and in L\'{e}vy walk, it takes a {\em constant velocity}. L\'{e}vy flight and L\'{e}vy walk can show the same pattern of traces but their time durations taken to have such traces are essentially different. Intuitively, L\'{e}vy flight can be easily slotted while L\'{e}vy walk is not.

Unfortunately, understanding tradeoffs between throughput and delay under L\'{e}vy mobility models is technically very challenging and underexplored. Unlike the other random walk models permitting mathematical tractability, the L\'{e}vy process is not very well understood mathematically despite significant studies on L\'{e}vy process in mathematics and physics. Thus, the conventional techniques \cite{SL_Sharma07, SL21Lin} used to study delay-capacity tradeoffs cannot be applied to L\'{e}vy mobility models, especially to L\'{e}vy walk which has high spatio-temporal correlation. In more specific, since L\'{e}vy walk is not eligible for discretization for Markovian analysis, its mathematical characteristics such as joint spatio-temporal probability density function (PDF) are hardly known. Due to such a difficulty, analyzing L\'{e}vy walk is generally considered to be very challenging.

Our main contribution is to analytically derive important tradeoffs between delay and capacity for both L\'{e}vy mobility models. An important point in this tradeoff is the ``critical delay'' which is the minimum delay for a mobile network to obtain a larger throughput than $\Theta(1/\sqrt{n})$. Our technique involves (i) analyzing the joint spatio-temporal PDF of a time-varying location of a node and the diffusion equation of the node for L\'{e}vy flight and (ii) characterizing an embedded Markov process inherent in L\'{e}vy walk which is a semi-Markov process. Since a different value of $\alpha$ induces a different mobility pattern, it also induces a different critical delay. Below we summarize our main results.

\smallskip
  \begin{center}
    {\small
  \begin{tabular}{|c|c|c|} \hline
    Mobility & $\alpha$ & Critical Delay \\ \hline \hline
    L\'{e}vy walk	& $\alpha \in (0,1)$ 	& $\Theta (\sqrt{n})$ 		\\ \hline
    				& $\alpha \in [1,2]$ 	& $\Theta (n^{\alpha / 2})$ \\ \hline \hline
    L\'{e}vy flight	& $\alpha \in (0,2]$	& $\Theta (n^{\alpha /2})$  \\ \hline
  \end{tabular}
}
  \end{center}
\smallskip

Given that many human mobility traces are shown to have values of $\alpha$ between 0.53 and 1.81~\cite{rhee:levymobility}, according to our results, mobile networks assisted by human mobility have critical delays between $\Theta (n^{0.27})$ and $\Theta(n^{0.91})$. Note that our results give much more detailed prediction of the critical delay for such mobile networks depending on $\alpha$ while Brownian motion and random waypoint always show $\Theta(n)$ and $\Theta(n^{0.5})$ for their critical delays~\cite{SL_Sharma07}.

The rest of the paper is organized as follows. We first overview a list of related work in Section~\ref{sec:related} and introduce our system model in Section~\ref{sec:model}. More details of L\'{e}vy mobility model parameterized by $\alpha$ are described in Section~\ref{sec:levymobility}, and the critical delays under L\'{e}vy flight and L\'{e}vy walk are investigated in Sections~\ref{sec:analysis:LF} and~\ref{sec:analysis:LW}, respectively. Finally, we provide a high level interpretation of our main results in Section~\ref{sec:discussion} and concluding remarks in Section~\ref{sec:conclusion}.

\section{Related Work}\label{sec:related}
Gupta and Kumar~\cite{SL1Gupta} showed that the per-node throughput of random wireless networks with $n$ static nodes scales as a function of $O(1/\sqrt{n})$ and proposed a scheme achieving $\Theta(1/\sqrt{n \log{n}})$. The result for static wireless networks was later enhanced to $\Theta (1/\sqrt{n})$ by exercising individual power control~\cite{SL4Franceschetti,SL5Ashish}. Grossglauser and Tse~\cite{grossglauser:mobility} proved that a constant per-node throughput is achievable by using mobility when the nodes follow ergodic and stationary mobility models. This result disproved the conventional belief that node mobility can negatively impact network capacity as it causes connectivity breakup and channel quality degradation.

Many follow-up studies~\cite{SL8Bansal, SL12Perevalov,SL13Tsingos,SL14Neely,SL17Toumpis,SL18Sharma,SL20Lin} have been devoted to understand, characterize and exploit the tradeoffs between throughput and delay. Especially, the delay required to obtain the constant throughput $\Theta (1)$ has been later studied under various mobility models~\cite{SL14Neely,SL16Neely,SL18Sharma,SL20Lin}. The studies provided that the delay to obtain $\Theta (1)$ of per-node throughput becomes $\Theta(n)$ for most mobility models such as i.i.d. mobility, random direction, random waypoint and Brownian motion models.

Another interesting question that has attracted researchers is what should be the minimum delay to achieve asymptotically higher throughput than $\Theta(1/\sqrt{n})$, the per-node throughput of static networks. This has been studied under the notion of critical delay~\cite{SL21Lin,SL_Sharma07} for two families of random mobility models: {\it hybrid random walk} and {\it random direction}. The hybrid random walk model splits the network of size 1 with $n^{2\beta}$ cells and further splits a cell into $n^{1-2\beta}$ subcells for $\beta\in[0,1/2]$. Then, a node moves to a random subcell of an adjacent cell in every unit time slot. In this model, i.i.d. mobility corresponds to $\beta = 0$ and random walk mobility corresponds to $\beta = 1/2$. For any $\beta\in[0,1/2]$, critical delay is proved to be $\Theta (n^{2\beta})$. The random direction model chooses a random direction within $[0, 2\pi]$ and moves to the selected direction with a distance of $n^{-\gamma}$ with a velocity $n^{-1/2}$ for $\gamma\in[0,1/2]$. In this model, random waypoint and Brownian motion are represented with $\gamma = 0$ and $\gamma =1/2,$ respectively. The critical delay is proved to be $\Theta (n^{1/2 + \gamma})$.

\section{Model Description}\label{sec:model}

\subsection{System Model}
We consider a wireless mobile network indexed by $n$, where in the $n$-th network, $n$ nodes are distributed uniformly on a completely wrapped-around square $\mathcal{S}(n)$ whose width and height scale as $\sqrt{n}$ and the density is fixed to 1 with increasing~$n.$\footnote{This model is often referred to as an extended network model. In another model, called a unit network model, the network area is fixed to 1 and the density increases as $n$ while the spacing and velocity of nodes scale as $1/\sqrt{n}$.} Without loss of generality, we set the width and the height of the square $\mathcal{S}(n)$ as $\sqrt{n}$. We assume that all nodes are homogeneous in that each node generates data with the same intensity to a per-source destination. The packet generation process at each node is assumed to be independent of node mobility.

A source-to-destination packet can be delivered by either direct one-hop transmission or over multiple hops, say $k$ hops, using relay nodes. We call it $k$-hop relay transmission. We assume that all nodes can serve as relay nodes for other source nodes and the nodes serving as relay nodes can only forward packets rather than replicating packets (not to overproduce the same packets in the network).

To model interference in wireless networks, we use the protocol model as in \cite{SL1Gupta}, under which nodes transmit packet successfully at a constant rate $W$ bits/sec, if and only if the following is met: let $\bm{X}_i(t)\,(\in\mathbb{R}^2)$ denote the location of node~$i\,(i=1,\ldots,n)$ at time~$t\,(\ge 0)$. For a transmitter~$i$, a receiver~$j$ and every other node~$k \neq i,j$ transmitting simultaneously,
\begin{align*}
d(\bm{X}_k(t), \bm{X}_j(t)) \ge (1+\Delta)\,d(\bm{X}_i(t), \bm{X}_j(t)),
\end{align*}
where $d(\bm{x},\bm{y})$ denotes the Euclidean distance between locations $\bm{x},\bm{y}\in\mathbb{R}^2$, and $\Delta$ is some positive number.

A packet can be delivered through a scheduling scheme which consists of {\em replication} or {\em forwarding}. We assume that only source nodes replicate packets and all other relay nodes forward them. As the names imply, replication copies a packet and the packet transmitter keeps the packet, whereas in forwarding the transmitter does not keep the original packet after successful transmission. This selective replication and forwarding depending on the node type are often applied to suppress the overflow of redundant packets in the network. Packets are delivered in two ways: {\it neighbor capture} and {\it multi-hop capture}. In neighbor capture, using mobility, relay or source nodes are located within the communication range of the destination. In the multi-hop capture, a source establishes a multi-hop path to the destination and delivers the packets over the path. We assume a fluid packet model~\cite{SL22Gamal} so that the delivery can occur immediately even in the case of multi-hop capture because the transmission delay is negligible compared to the delay from node mobility. We denote by $\Pi$ the class of all scheduling schemes conforming the descriptions above.

\subsection{Performance Metrics}
The primary performance metric in many networking systems is per-node throughput measured by the long-term average of received packets aggregated over nodes:
\smallskip
\begin{definition}[Per-node throughput] Let $\lambda_\pi(n)$ denote the per-node throughput in the $n$-th network under a scheduling scheme $\pi\in\Pi$. It is then given by
\begin{align*}
\lambda_{\pi}(n) \triangleq \liminf_{t \rightarrow \infty} \frac{1}{n} \sum_{i=1}^{n} \frac{\lambda_{\pi:i} (t)}{t},
\end{align*}
where $\lambda_{\pi:i}(t)$ is the total number of bits received at a destination node $i$ up to time $t$ under $\pi$.\footnote{For simplicity, we omit the subscript $\pi$ in $\lambda_\pi(n)$ unless confusion arises.}
\end{definition}
\smallskip

Another important metric is average delay:
\smallskip
\begin{definition}[Average delay] Let $D_\pi(n)$ denote the average delay in the $n$-th network under a scheduling scheme $\pi\in\Pi$. It is then given by
\begin{align*}
D_{\pi}(n) \triangleq \lim_{k \rightarrow \infty} \frac{1}{n} \sum_{i=1}^{n} \frac{1}{k} \sum_{j=1}^{k} D_{\pi:(i,j)},
\end{align*}
where $D_{\pi:(i,j)}$ is the individual packet delay that a packet $j$ experiences to arrive at a destination node $i$ from its source node under $\pi$.
\end{definition}
\smallskip

We give special attention to the notion of critical delay, first introduced in \cite{SL_Sharma07}:
\smallskip
\begin{definition}[Critical delay] The critical delay in the $n$-th network, denoted by $C_{\Pi}(n)$, is the minimum average delay that must be tolerated under a given mobility model to achieve a per-node throughput of $\omega(1/\sqrt{n})$, i.e.,
\begin{align*}
C_{\Pi}(n) \triangleq \inf_{ \{\pi \in \Pi:\lambda_{\pi}(n) = \omega(1/ \sqrt{n})\}} D_{\pi}(n).
\end{align*}
\end{definition}
\smallskip

Per-node throughput $\Theta(1/\sqrt{n})$ is achievable by a scheduling scheme in static multi-hop networks \cite{SL1Gupta}. Since node mobility can increase per-node throughput at the cost of larger delay, the critical delay quantifies the amount of delay that a network should sacrifice to achieve the guaranteed ``baseline'' per-node throughput. It can be used as a simple, yet useful metric for a mobility model, representing how sensitive the delay is to increase per-node throughput.

Computing critical delay consists of multiple steps. We start by following the initial step in~\cite{SL21Lin,SL_Sharma07} which connects critical delay to the first exit time. Let $\mathcal{D}(n)$ denote a disc within the square $\mathcal{S}(n)$ whose radius scales as $\Theta(\sqrt{n})$. Critical delay can simply be regarded as {\em the maximum time duration that a node cannot exit from the disc $\mathcal{D}(n)$ with probability approaching~1 as $n$ goes to $\infty$.} In our extended network model, the average distance from a source node to a destination node is $\Theta(\sqrt{n})$ when they are uniformly distributed on $\mathcal{S}(n)$. Therefore, if nodes travel up to a distance $O(\sqrt{n})$, for a certain time duration, the distance between a source or a relay and a destination still remains $\Theta(\sqrt{n})$ on average which results in $O(1/\sqrt{n})$ per-node throughput (see Lemma~\ref{lemma:throughput}). Thus, it is obvious that a network aiming at obtaining $\omega(1/ \sqrt{n})$ per-node throughput must allow a delay which is no less than the maximum time duration that the first exit of a node from the disc $\mathcal{D}(n)$ does not occur with probability approaching~1. This insight can be formally described with the notion of the first exit time:

\smallskip
\begin{definition}[First exit time]\label{def:fet} Let $\bm{X}_i (0) = \bm{x}$. The first exit time for a disc of a radius~$r$, denoted by $T(r)$, is defined as
\begin{align*}
T(r) \triangleq \inf \{t\geq 0 : \bm{X}_i (t) \notin B(\bm{x},r) \},
\end{align*}
where $B(\bm{x},r)$ denotes the set of points $\bm{y}$ in $\mathcal{S}(n)$ such that $d(\bm{x},\bm{y}) \leq r$.
\end{definition}
\smallskip
Without loss of generality, we set the radius of the disc $\mathcal{D}(n)$ as $c_d\sqrt{n}$ where $c_d$ is a constant in the range $(0, 1/2)$. Then, critical delay $C_{\Pi}(n)$ can be obtained by
\begin{align*}
C_{\Pi}(n) = \sup \left\{t(n) : \lim_{n \rightarrow \infty} \pr\{T (c_d\sqrt{n}) > t(n)\} = 1 \right\}.
\end{align*}

\smallskip
\begin{lemma}[\cite{SL21Lin,SL1Gupta}]\label{lemma:throughput}
Suppose that on average each packet is relayed over a total distance no less than $\Theta(\sqrt{n})$ in an extended network model. Then, the per-node throughput $\lambda(n)$ scales as $O(1/\sqrt{n})$.
\end{lemma} 
\section{Mobility Models: L\'{e}vy Flight and L\'{e}vy Walk}\label{sec:levymobility}
In this section, we formally define {\em L\'{e}vy mobility model: L\'{e}vy flight and L\'{e}vy walk}.

L\'{e}vy flight and L\'{e}vy walk processes are treated separately in the literature~\cite{RW01Drysdale, RW05Metzler, Chechkin2006}. L\'{e}vy flight takes a {\em constant time} for any step irrespective of its step size, whereas L\'{e}vy walk takes a {\em constant velocity} for every step. Thus, in L\'{e}vy walk, the time taken for each step is proportional to the step size. The distinction between L\'{e}vy flight and L\'{e}vy walk is often made based on the speeds of their actual processes. L\'{e}vy flight is a ``fast'' mobility model which can reach its next destination in a constant time no matter how far it is. In a similar context, L\'{e}vy walk falls under a ``slow'' mobility model. An experimental velocity model suggested as a function of step size in~\cite{rhee:levymobility} verifies that a human mobility lies in between L\'{e}vy flight and L\'{e}vy walk. For convenience, we use L\'{e}vy mobility model to indicate both of L\'{e}vy flight and L\'{e}vy walk, unless explicitly stated.

Let $Z$ be a random variable denoting the step size under L\'{e}vy mobility model. Then, $Z$ is generated from a random variable $\dot{Z}$ having the L\'{e}vy $\alpha$-stable distribution~\cite{nolan:2012} by the relation $Z = |\dot{Z}|$. The PDF of $\dot{Z}$ is give by
\begin{align}\label{eq:exact_pdf_flight_length}
f_{\dot{Z}}(z) = \frac{1}{2\pi} \int_{-\infty}^{\infty} e^{-izt}\varphi_{\dot{Z}}(t) \, \text{d}t,
\end{align}
where $\varphi_{\dot{Z}}(t)\deq \E[e^{it\dot{Z}}]$ is the characteristic function of~$\dot{Z}$ and is given by $\varphi_{\dot{Z}}(t)= e^{-|ct|^\alpha}$. Here, $|c|> 0$ is a scale factor which is a measure of the width of the distribution, and $\alpha \in (0,2]$ is a distribution parameter and specifies the shape (i.e., heavytail-ness) of the distribution. The step size~$Z$ for $\alpha \in (0,1)$ has infinite mean and variance, while $Z$ for $\alpha \in [1,2)$ has finite mean but infinite variance. For $\alpha= 2$, the L\'{e}vy $\alpha$-stable distribution reduces to a Gaussian distribution with zero mean and variance $\sigma^2 = 2c^2$, and consequently the step size $Z$ has finite mean and variance.

Due to the complex form of the distribution, the L\'{e}vy $\alpha$-stable distribution for $\alpha \in (0,2)$ is often given as a power-law type of asymptotic form, closely approximating the tail part of the distribution~\cite{nolan:2012}:
\begin{align}\label{eq:app_pdf_flight_length}
f_{\dot{Z}}(z) \sim \frac{1}{|z|^{1+\alpha}}.
\end{align}
For mathematical tractability, in our analysis we use the asymptotic form~(\ref{eq:app_pdf_flight_length}) instead of the exact form~(\ref{eq:exact_pdf_flight_length}) for $\alpha\in(0,2)$ while using the exact form~(\ref{eq:exact_pdf_flight_length}) for $\alpha=2$. The form (\ref{eq:app_pdf_flight_length}) is known to closely approximate~(\ref{eq:exact_pdf_flight_length}) and several papers in mathematics and physics, e.g.,~\cite{RW01Drysdale,PhysRevE.73.057102}, analyze L\'{e}vy mobility using form~(\ref{eq:app_pdf_flight_length}). For the range of $Z$, since we use the extended network model, the step size~$Z$ is assumed to have a lower bound at 1 and an upper bound at $\sqrt{n}$, i.e., $\pr\left\{1 \le Z \le \sqrt{n}\right\} = 1$.\footnote{The bounds are chosen equivalently to the lower bound at $1/\sqrt{n}$ and the upper bound at $1$ for the step size in the unit network model~\cite{SL_Sharma07}.} Thus, the complementary cumulative distribution function (CCDF) of $Z$ becomes $\pr\{Z>z\} = 1$ for $z <1$ and $\pr\{Z>z\} = 0$ for $z \ge \sqrt{n}$. For $z\in[1, \sqrt{n})$, we have
\begin{align}\label{eq:model:ccdf}
\pr\{Z\!>\!z\}
&\!=\!
\begin{cases}
c(n)\cdot \!\big(\frac{1}{z^\alpha} \!-\! \frac{1}{(\sqrt{n})^\alpha}\big) & \text{for } \alpha \in (0,2), \\
c(n)\cdot \! \big(\mathrm{erf}(\frac{\sqrt{n}}{\sqrt{2}\sigma})\! -\! \mathrm{erf}(\frac{z}{\sqrt{2}\sigma})\big) & \text{for } \alpha = 2.
\end{cases}
\end{align}
Here, $\mathrm{erf}(\cdot)$ is the error function defined as $\mathrm{erf}(x)\deq \frac{2}{\sqrt{\pi}}\int_{0}^{x} \exp(-t^2)\, \text{d}t$, and $c(n)$ is defined as\footnote{To be precise, $c(n)$ is also a function of $\alpha$, i.e., $c(n) = c(n,\alpha)$. Since we focus on scaling properties with respect to $n$ for a fixed $\alpha$, we omit the argument $\alpha$ in $c(n,\alpha)$ for notational simplicity. By the same reason, in the rest of the paper, we emphasize only $n$ in all variables that depend on both $n$ and $\alpha$.}
\begin{align*}
c(n) \deq
\begin{cases}
(1-\frac{1}{(\sqrt{n})^\alpha})^{-1} & \text{ for } \alpha\in(0,2), \\
(\mathrm{erf}(\frac{\sqrt{n}}{\sqrt{2}\sigma})- \mathrm{erf}(\frac{1}{\sqrt{2}\sigma}))^{-1} & \text{ for } \alpha = 2.
\end{cases}
\end{align*}
Note that as $n$ goes to $\infty$, the CCDF $\pr\{Z>z\}$ for $z \ge 1$ goes to $1/z^{\alpha}$ for $\alpha\in(0,2)$.

In our analysis, we use the following assumptions on the L\'{e}vy mobility model: (A1) the time taken for each step in the L\'{e}vy flight is set to 1, and (A2) the velocity taken for each step in L\'{e}vy walk is set to 1. Note that as long as these two metrics are constant, the scaling property of critical delay remains the same, which justifies our assumptions.

\section{Critical Delay Analysis for L\'{e}vy Flight}\label{sec:analysis:LF}
In this section, we will show that the critical delay $C_{\Pi}(n)$ under L\'{e}vy flight with a distribution parameter $\alpha\in(0,2]$ scales as $\Theta(n^{\frac{\alpha}{2}})$ (Theorem~\ref{thm:critical_delay_LF}). In Section~\ref{sec:analysis:LF:tech}, we explain technical challenges and our approach for proving Theorem~\ref{thm:critical_delay_LF}. In Section~\ref{sec:analysis:LF:ana}, we prove Theorem~\ref{thm:critical_delay_LF} by showing that the upper bound on $C_{\Pi}(n)$ scales as $O(n^{\frac{\alpha}{2}})$ (Lemma~\ref{lemma:scaling_property_1_LF}), and that the lower bound on $C_{\Pi}(n)$ scales as $\Omega(n^{\frac{\alpha}{2}})$ (Lemma~\ref{lemma:scaling_property_2_LF}).

\subsection{Technical Approach}\label{sec:analysis:LF:tech}
We begin with deriving a relation between the first exit time of a 2-dimensional random process and the one for its 1-dimensional projected process. We then describe trapping phenomenon in a diffusion process that have a direct connection to the first exit time of a 1-dimensional random process.

It is clear from Definition~\ref{def:fet} that the statistical proprieties of the first exit time do not depend on the choice of node index~$i$. Thus, we omit the node index~$i$ in the rest of the paper. Denote $\bm{X}(t) = (X_x(t), X_y(t))$ and consider the projected processes $\{X_{x}(t)\}_{t \ge 0}$ and $\{X_{y}(t)\}_{t \ge 0}$ onto $x$-axis and $y$-axis, respectively.
We define for the projected processes the first exit time similarly to Definition~\ref{def:fet}:
\begin{align*}
T_{x}(r) &\triangleq \inf \left\{t \ge 0 : |X_{x}(t) - X_{x}(0)| \ge r\right\}, \\
T_{y}(r) &\triangleq \inf \left\{t \ge 0 : |X_{y}(t) - X_{y}(0)| \ge r\right\}.
\end{align*}
Since the event $\{|X_x(t) - X_x(0)| \ge r\}$ implies the event $\{d(\bm{X}(t), \bm{X}(0)) \ge
r\}$, we obtain
\begin{align}\label{eq:from_2D_to_1D_mod}
\pr\{T_{x}(r) \le t\} \le \pr\{T(r) \le t\}.
\end{align}
In addition, it is clear that
\begin{align}\label{eq:from_2D_to_1D}
\pr \{T(r) \le t \}
&\le \pr \{T_{x}(r/\sqrt{2}) \le t \text{ or } T_{y}(r/\sqrt{2}) \le t \} \nonumber \\
&\le 2 \, \pr \{T_{x}(r/\sqrt{2}) \le t \},
\end{align}
where the second inequality comes from the union bound and the symmetry of node motion.
Combining~(\ref{eq:from_2D_to_1D_mod}) and~(\ref{eq:from_2D_to_1D}), we have for all $t \ge 0$,
\begin{align}\label{eq:lower_and_upper_prob}
\pr\{T_{x}(r) \le t\} &\le \pr\{T(r) \le t\} \le 2 \, \pr \{T_{x}(r/\sqrt{2}) \le t \}.
\end{align}
Our technical approach is mainly based on (\ref{eq:lower_and_upper_prob}), and is to bound the
first exit time distribution for 2-dimensional L\'{e}vy flight by the one for the corresponding
1-dimensional projected process $\{X_x(t)\}_{t\ge 0}$. We henceforth study the first exit time
distribution for the process $\{X_x(t)\}_{t\ge 0}$.

The first exit time analysis for 1-dimensional random processes has been intensively studied in
physics and mathematics, e.g.,~\cite{RW09Redner}. Specifically, trapping phenomenon (of a
diffusing particle) in physics and its related theories have a direct connection to our first
exit time problem as explained in the following: consider a particle that diffuses in a finite
interval $[0, 2r]\,(\subset\mathbb{R})$ having trapping boundaries at $x=0, 2r$. Let
$L(t)\,(\in\mathbb{R})$ be a random variable denoting the location of the particle at time~$t$.
The particle is assumed to be initially located at $L(0) = r$, and eventually it is trapped at
either of both boundaries with probability~1. Upon the particle is trapped, it disappears in the
interval. We call the state of the particle \emph{survival state} until the particle is trapped
and disappears. By convention, we let $L(t)\deq \varnothing$ if the particle is not in survival
state at time $t$. If we assume $X_x(0) = L(0) \, (= r)$, then $X_x(t)$ and $L(t)$ for $t>0$ are
related as follows:
\begin{align}\label{eq:L(t) and X_x(t)}
\begin{split}
L(t) \ed
\begin{cases}
X_x(t) & \text{ if } t < T_x(r), \\
\varnothing & \text{ if } t \ge T_x(r),
\end{cases}
\end{split}
\end{align}
where $\ed$ denotes ``equal in distribution". Hence, we have from~(\ref{eq:L(t) and X_x(t)})
that
\begin{align}\label{eq:L(t) and T_x(r)}
\pr\{T_x(r) \le t\} = \pr\{L(t) = \varnothing\}.
\end{align}
That is, the survival time of a particle in the trapping model has the same distribution as the
first exit time $T_x(r)$ of a node under L\'{e}vy flight.

The technical approach for analyzing the critical delay in the literature is as follows. In the
case of Brownian motion, there are two general techniques in studying the critical delay. One is
to discretize mobility and then apply a Markovian analysis~\cite{SL_Sharma07}. The other is to
use a continuous mobility model and solve a diffusion equation to obtain a joint spatio-temporal
PDF of a time-varying location of a node~\cite{SL21Lin}. The latter enables one to obtain the
distribution of $L(t)$ whose spatial derivative is often referred to as \emph{occupation
probability}.\footnote{The occupation probability in a trapping model corresponds to the joint
spatio-temporal PDF in a random walk model. The mathematical definition and the distinction
between the occupation probability and the joint spatio-temporal PDF will be given in
Section~\ref{sec:analysis:LF:ana}.} The occupation probability of Brownian motion can be decomposed
to find the components constituting it. From this decomposition process, we find that there is a dominating term which characterizes the limiting behavior of the first exit time distribution.

In the case of L\'{e}vy flight, the joint spatio-temporal PDF has a similar form to that of
Brownian motion. In addition, the occupation probabilities and the first exit time distributions for Brownian motion and L\'{e}vy flight have similar structures in the aspect of the dominating terms. Hence, by identifying and characterizing the dominating term for L\'{e}vy flight, we can obtain the critical delay under L\'{e}vy flight.

%
%
\subsection{Analysis}\label{sec:analysis:LF:ana}
In this subsection, we provide the detailed result for the critical delay under L\'{e}vy flight. Our main result is derived by following three steps: (i) the occupation probability is obtained
from the solution of a differential equation that governs the movement of a particle. (ii) From
the occupation probability, we obtain the survival probability (which will be defined later),
which in turn yields the first exit time distribution. (iii) By investigating the limiting
behavior of the first exit time distribution, we can finally obtain the order of the critical
delay.

\smallskip
\textit{Step 1:} Let $P(x,t) \deq \frac{\text{d}}{\text{d}x} \pr\{L(t) \le x\}$. Intuitively,
$P(x,t)$ represents probability that the particle is located at~$x$ at time~$t$. We call $P(x,t)$
the \emph{occupation probability}, and it has the following properties:
\begin{itemize}
\item (P1) $\lim_{t\to\infty} P(x,t) = 0~\forall x \in \mathbb{R}.$
\item (P2) $\int_{0}^{2r} P(x,0)\, \text{d}x = \pr\{L(0)=r\} = 1$.
\item (P3) $\int_{0}^{2r} P(x,t)\, \text{d}x \le 1~\forall t >0$.
\item (P4) $P(0,t) = P(2r, t) = 0~\forall t \ge 0$.
\item (P5) Since $P(x,0)$ is a PDF having a support~$\{r\}$, we have $P(x,0) = \delta_{x,r}$,
    where $\delta_{x_1,x_2}$ denotes the Kronecker delta which is defined to be 1 if
    $x_1=x_2$ and 0 otherwise.
\end{itemize}

To be precise, $P(x,t)$ for $t>0$ could not be a PDF due to (P3). However, the function obtained
by normalizing $P(x,t)$ with the integral $\int_{0}^{2r} P(x,t)\, \text{d}x$, denoted by
$\bar{P}(x,t)$, becomes a PDF for a finite time $t$. We call $\bar{P}(x,t)$ the \emph{joint
spatio-temporal PDF} at location $x$ and time $t$.

In the first step, we obtain the occupation probability $P(x,t)$ for the process $\{L(t)\}_{t \ge
0}$. For this, we need to characterize the associated 1-dimensional process $\{X_x(t)\}_{t\ge
0}$. We first consider the case of $\alpha\in(0,2)$ and summarize the result in the following lemma.

\smallskip
\begin{lemma}\label{lemma:1D_projection_LF}
Suppose that $\{\bm{X}(t)\}_{t \ge 0}$ is 2-dimensional L\'{e}vy flight with a distribution
parameter $\alpha\in(0,2)$. Then, as $n$ goes to $\infty$, the projected process onto $x$-axis
$\{X_x(t)\}_{t \ge 0}$ approaches to 1-dimensional L\'{e}vy flight having the same distribution
parameter~$\alpha$. It holds for the process $\{X_y(t)\}_{t\ge 0}$.

\smallskip
\noindent\textit{Proof:} Let $Z_i$ and $\theta_i\,(i=1,2,\ldots)$ be random variables denoting
the $i$-th step size and direction of the process $\{\bm{X}(t)\}_{t\ge 0}$, respectively. Then,
$\bm{X}(t)$ for $t=1,2,\ldots$ can be expressed as
\begin{align}\label{eq:LF:location}
\bm{X}(t) &= (X_x(t), X_y(t)) \nonumber \\
&= \bm{X}(0) + \left(\sum_{i=1}^{t} Z_i \cos \theta_i,\, \sum_{i=1}^{t} Z_i \sin
\theta_i\right).
\end{align}
We will show that, as $n$ goes to $\infty$, arbitrary step size of the projected processes (i.e.,
$Z_i |\cos \theta_i|$ and $Z_i |\sin \theta_i|$) has a power-law type CCDF with an exponent
$\alpha$, i.e., for $z \ge 1$,
\begin{align}\label{eq:LF:ccdf}
\begin{split}
\lim_{n\to\infty}\pr\{Z_i |\cos \theta_i| > z\}
&= \lim_{n\to\infty}\pr\{Z_i |\sin \theta_i| > z\} \\
&= \frac{c^\star}{z^{\alpha}},
\end{split}
\end{align}
where $c^\star \deq \frac{2}{\pi}\int_{0}^{\frac{\pi}{2}}(\cos\vartheta)^\alpha
\text{d}\vartheta$. Since the projected processes take a constant time for every step
irrespective of step size, the property in (\ref{eq:LF:ccdf}) proves the lemma.

Now we prove (\ref{eq:LF:ccdf}). By conditioning on the values of the random
variable~$\theta_i\sim \text{Uniform}[0,2 \pi]$, we can rewrite the CCDF of $Z_i |\cos \theta_i|$
as
\begin{align}\label{eq:projected_distribution}
\pr\{Z_i |\cos \theta_i| > z\}
&= \int_{0}^{2\pi} \pr\{Z_i |\cos \theta_i| > z \, | \, \theta_i = \vartheta\}
\text{d}F_{\theta_{i}}(\vartheta) \nonumber \\
&= \frac{1}{2\pi} \int_{0}^{2\pi} \pr\{Z_i |\cos \vartheta| > z\}  \text{d}\vartheta \nonumber
\\
&= \frac{2}{\pi} \int_{0}^{\frac{\pi}{2}} \pr\{Z_i \cos \vartheta > z\} \text{d}\vartheta,
\end{align}
where the last two equalities come from the independence of the random variables $Z_i$ and
$\theta_i$, and the symmetry of the function $|\cos\vartheta|$, respectively. Using
(\ref{eq:model:ccdf}), the probability $\pr\{Z_i \cos \vartheta > z\}$ in
(\ref{eq:projected_distribution}) can be obtained for $\vartheta \in [0,\frac{\pi}{2}]$ as
\begin{align*}
\lefteqn{\pr\{Z_i \cos \vartheta > z\}} \\
&=
\begin{cases}
c(n)\cdot\big((\frac{\cos\vartheta}{z})^{\alpha} -(\frac{1}{\sqrt{n}})^{\alpha}\big) & \text{for
} \vartheta \in [0, \cos^{-1}(\frac{z}{\sqrt{n}})), \\
0 & \text{for } \vartheta \in [\cos^{-1}(\frac{z}{\sqrt{n}}), \frac{\pi}{2}].
\end{cases}
\end{align*}
Hence, the CCDF $\pr\{Z_i |\cos \theta_i| > z\}$ is given by
\begin{align}\label{eq:LF:ccdf_2}
\begin{split}
\pr\{Z_i |\cos \theta_i| > z\} & = \frac{2 c(n)}{\pi z^\alpha}
\int_{0}^{\cos^{-1}(\frac{z}{\sqrt{n}})}  (\cos\vartheta)^\alpha \text{d}\vartheta \\
& \qquad -\frac{2
c(n)}{\pi (\sqrt{n})^\alpha}  \cos^{-1}\left(\frac{z}{\sqrt{n}}\right).
\end{split}
\end{align}
Noting $\lim_{n\to\infty}c(n)=1$ and $\lim_{n\to\infty}\cos^{-1}\left(\frac{z}{\sqrt{n}}\right) =
\frac{\pi}{2}$, we have from (\ref{eq:LF:ccdf_2}) that
\begin{align*}
\lim_{n\to\infty}\pr\{Z_i |\cos \theta_i| > z\} &= \frac{2}{\pi z^\alpha}
\int_{0}^{\frac{\pi}{2}} (\cos\vartheta)^\alpha \text{d}\vartheta  = \frac{c^\star}{z^\alpha}.
\end{align*}
Since $|\sin\theta_i| \ed |\cos\theta_i|$ for $\theta_i\sim \text{Uniform}[0, 2\pi]$, we have
\begin{align*}
    \pr\{Z_i |\sin \theta_i| > z\}
    &= \pr\{Z_i |\cos \theta_i| > z\},
\end{align*}
which completes the proof. \hfill$ \square$
\end{lemma}
\smallskip

Motivated by Lemma~\ref{lemma:1D_projection_LF} and (\ref{eq:L(t) and X_x(t)}), we study the occupation probability for 1-dimensional L\'{e}vy flight with $\alpha\in(0,2)$ in a finite interval $[0, 2r]$ having trapping boundaries. For mathematical tractability, our study in this subsection assumes continuous limit where the scale factor~$|c|$ in~(\ref{eq:exact_pdf_flight_length}) approaches to zero. Then, the occupation probability $P(x,t)$ for $\alpha\in(0,2)$ is governed by the following fractional Fokker-Planck equation~\cite[Eq.~(22)]{Chechkin2006},~\cite[Eq.~(28)]{RW04Gitterman}:
\begin{align}\label{eq:governing_equation_LF}
\frac{\partial P(x,t)}{\partial t} = F \frac{\partial^{\alpha} P(x,t)}{\partial |x|^{\alpha}},
\end{align}
where $F \,(= F_{\alpha}>0)$ is a generalized diffusion coefficient and $\frac{\partial^{\alpha}}{\partial |x|^{\alpha}}$ is the Riesz-Feller derivative of fractional order $\alpha$~\cite{Podlubn1999}. We next consider the case of $\alpha=2$. In this case, as the scale factor~$|c|$ approaches to zero, the 2-dimensional L\'{e}vy flight converges to a Wiener process which mathematically models a continuous movement of Brownian motion. Since 1-dimensional projected process of 2-dimensional Brownian motion is also Brownian motion~\cite{SL21Lin}, the occupation probability for $\alpha=2$ is governed by the normal diffusion equation where the spatial derivative of order $\alpha$ with $\alpha\in(0,2)$ in (\ref{eq:governing_equation_LF}) is replaced by the second order derivative with $\alpha=2$~\cite{RW09Redner}. Therefore, with continuous limit, the occupation probability $P(x,t)$ for $\alpha\in(0,2]$ can be described by the differential equation in~(\ref{eq:governing_equation_LF}). Through Appendix~\ref{sec:appendix:A}, we show that the order of the critical delay under L\'{e}vy flight does not change with continuous limit.

Applying the standard method of separation of variables gives the
solution of~(\ref{eq:governing_equation_LF}) as follows:
\begin{align}\label{eq:occupation_prob_LF}
P(x,t) = \sum_{i=1}^{\infty} h_{i} \psi_{i}(x) \exp \left( \lambda_{i}F t \right).
\end{align}
Here, $h_{i} \, (i=1,2,\ldots)$ are determined from the initial condition $P(x, 0) =
\delta_{x,r}$ (as shown in (P5)) and are given by $h_{i} = \psi_{i}(r)$. The functions
$\psi_{i}(x)$ and the constants $\lambda_{i}$ can be obtained from the solutions of the
problem $\mathfrak{D}[\psi_{i}(x)]= \lambda_{i} \psi_{i}(x)$ for the operator
$\mathfrak{D}\triangleq\frac{\text{d}^{\alpha}}{\text{d} |x|^{\alpha}}$, and are called
eigenfunctions and eigenvalues of~$\mathfrak{D}$, respectively. Without loss of generality, we
assume that $\lambda_{i}$ are arranged as $|\lambda_{1}| < |\lambda_{2}| < \cdots$.

\smallskip
\textit{Step 2:} Let $S(t) \deq \pr\{L(t) \neq \varnothing\}$. Intuitively, $S(t)$ represents
probability that the particle has not hit any trapping boundary by time $t$. We call $S(t)$ the
\emph{survival probability}. The survival probability can be obtained from the occupation
probability $P(x,t)$ by $S(t) = \int_{0}^{2r} P(x,t)\, \text{d}x$. Thus,
from~(\ref{eq:occupation_prob_LF}), the survival probability is given by
\begin{align}\label{eq:survival prob}
S(t) = \sum_{i=1}^{\infty}  \psi_{i}(r) \int_{0}^{2r}\psi_{i}(x)\, \text{d}x \exp \left( \lambda_{i}F t \right).
\end{align}
The first exit time distribution $\pr\{T_x(r) \le t\}$ can be obtained from the survival
probability $S(t)$ through the following relation:
\begin{align}\label{eq:FET_form}
\pr\{T_x(r) \le t\} &= \pr\{L(t) = \varnothing\} = 1 - S(t).
\end{align}
Here, the first equality comes from (\ref{eq:L(t) and T_x(r)}) and the second equality comes from
the definition of $S(t)$. By combining (\ref{eq:survival prob}) and (\ref{eq:FET_form}), we obtain the first exit time distribution in terms of the eigenfunctions $\psi_{i}(x)$ and the eigenvalues $\lambda_i$ as follows:
\begin{align}\label{eq:FET_formula}
\pr\{T_x(r) \le t\} &= 1 - \sum_{i=1}^{\infty} \! \psi_{i}(r) \! \int_{0}^{2r}\!\!\!\psi_{i}(x) \text{d}x \exp \left( \lambda_{i}F t \right).
\end{align}
For $\alpha=2$, the eigenfunctions and the eigenvalues in (\ref{eq:FET_formula}) can be obtained from the boundary
conditions $P(0,t) = P(2r,t) = 0~\forall t \ge 0$ (as shown in (P4)), and are given by
$\psi_{i}(x) = \sqrt{\frac{1}{r}} \sin \left(\frac{i \pi x}{2r} \right)$ and $\lambda_{i} =
-\left(\frac{i \pi}{2r}\right)^{2}$, respectively~\cite{RW09Redner}. For $\alpha\in(0,2)$,
Gitterman~\cite{RW04Gitterman} provided a solution of~(\ref{eq:governing_equation_LF}) whose eigenfunctions and eigenvalues are given by $\psi_{i}(x) = \sqrt{\frac{1}{r}} \sin \left(\frac{i \pi x}{2r} \right)$ and $\lambda_{i} = -\left(\frac{i \pi}{2r}\right)^{\alpha}$, respectively. Thus, under L\'{e}vy flight with $\alpha\in(0,2]$, the first exit time distribution can be
expressed as an infinite series of exponential functions as follows:
\begin{align}\label{eq:FPT_distribution_LF}
\pr\{T_x(r) \le t\} = 1 - \sum_{i=1}^{\infty} \eta_i \exp \left( - \frac{\rho_i}{r^\alpha} t
\right),
\end{align}
where $\eta_i \triangleq \frac{2 \{1 - \cos (i \pi)\}}{i \pi} \sin \left( \frac{i \pi}{2}
\right)$ and $\rho_i \triangleq F (\frac{i \pi}{2})^\alpha$.

As will be shown later in the proof of Lemmas~\ref{lemma:scaling_property_1_LF}
and~\ref{lemma:scaling_property_2_LF}, the smallest (i.e., dominant) decay rate in the
exponential functions in~(\ref{eq:occupation_prob_LF}) (i.e., $|\lambda_1|$) determines the
limiting behavior of the first exit time distribution. That is, the smallest decay rate
characterizes the critical delay under L\'{e}vy flight. The solutions
in~\cite{RW09Redner,RW04Gitterman} show that the dominant decay rate $|\lambda_{1}|$ scales as
$\Theta(r^{-\alpha})$ for $\alpha\in(0,2]$.

\smallskip
\textit{Step 3:} We are now ready to derive the main result of this subsection. By using the
closed-form expression for $\pr\{T_x(r) \le t\}$ in~(\ref{eq:FPT_distribution_LF}), we
investigate the order of the critical delay, stated in Lemmas~\ref{lemma:scaling_property_1_LF}
and~\ref{lemma:scaling_property_2_LF}.

%
%
\smallskip
\begin{lemma}[Upper bound for L\'{e}vy flight]\label{lemma:scaling_property_1_LF} Suppose that under L\'{e}vy flight with a distribution parameter $\alpha\in(0,2]$, the time $t\triangleq \hat{t}(n)$ in $\pr\{T(c_d \sqrt{n}) > t\}$ scales as $\Theta(n^{\frac{\alpha}{2} + \epsilon})$ for an arbitrary $\epsilon > 0$. Then, we have
\begin{align*}
\lim_{n\to\infty} \pr\{T(c_d\sqrt{n}) > \hat{t}(n)\} = 0,
\end{align*}
which shows that the critical delay $C_{\Pi}(n)$ under L\'{e}vy flight scales as $O(n^{\frac{\alpha}{2}})$.

\smallskip
\noindent\textit{Proof:} We will prove this lemma by showing that $\lim_{n \to \infty}
\pr\{T_x(c_d\sqrt{n}) \le \hat{t}(n)\} = 1$. Then, by substituting $r = c_d\sqrt{n}$ and $t=
\hat{t}(n)$ into~(\ref{eq:lower_and_upper_prob}) and taking a limit to $n$, we obtain
\begin{align*}
1 &= \lim_{n \to \infty} \pr\{T_x(c_d \sqrt{n}) \le \hat{t}(n)\} \\
&\le  \lim_{n \to \infty} \pr\{T(c_d \sqrt{n}) \le \hat{t}(n)\}.
\end{align*}
That is, $\lim_{n \to \infty} \pr\{T(c_d\sqrt{n}) \le \hat{t}(n)\} = 1$, or equivalently,
$\lim_{n \to \infty} \pr\{T(c_d\sqrt{n}) > \hat{t}(n)\} = 0$, which proves the lemma.

First, consider the case of $\alpha=2$. We substitute $r = c_d\sqrt{n}$ and $t= \hat{t}(n)$
into~(\ref{eq:FPT_distribution_LF}). Then, the series on the right-hand side
of~(\ref{eq:FPT_distribution_LF}) becomes a function of~$n$, and (for notational convenience) we
let
\begin{align*}
\pr\{T_x(c_d\sqrt{n}) \le \hat{t}(n)\} &= 1- \sum_{i=1}^{\infty} \eta_i \exp \left( -
\frac{\rho_i}{(c_d)^2 n} \hat{t}(n) \right)\\
&\deq 1- \hat{S}(n).
\end{align*}
We now need to take a limit to $\hat{S}(n)$. To validate the interchange of the order of limit
and summation, we will show that there exists a constant $\hat{n}\in\mathbb{N}$ such that the
infinite series $\hat{S}(n)$ converges uniformly on
$\hat{\mathcal{D}}\triangleq[\hat{n},\infty)$.\footnote{$\mathbb{N}$ denotes a set of positive
integers.} The uniform convergence will be shown by using the well-known Weierstrass $M$
test~\cite{RW06Marsden}.

Since $\hat{t}(n)=\Theta(n^{1 + \epsilon})$, there exist constants $\hat{n}\in\mathbb{N}$ and
$\hat{c}>0$ such that
\begin{align}\label{eq:BM:hat_t(n) expression}
\hat{t}(n) \ge \hat{c} n^{1+\epsilon} \quad \text{ for all } n\ge \hat{n}.
\end{align}
Let $\hat{m} \deq F(\pi/2 c_d)^2 \hat{c} (\hat{n})^{\epsilon}\,(>0)$. Then, the $i$-th function
of the series~$\hat{S}(n)$ is bounded by a constant $\hat{M}_i\triangleq \frac{4}{\pi}
\{\exp(-\hat{m})\}^i$ for all $n \ge \hat{n}$ as follows:
\begin{align*}
\bigg|\eta_i \exp \left( - \frac{\rho_i}{(c_d)^2 n}\hat{t}(n) \right)\bigg|
& \le \frac{4}{\pi} \exp \left( - \frac{\rho_i}{(c_d)^2 n} \hat{c} n^{1+\epsilon} \right) \\ &\le
\frac{4}{\pi} \exp\left( - \frac{F i (\pi)^2 }{4(c_d)^2} \hat{c} (\hat{n})^{\epsilon} \right) \\
&= \hat{M}_i.
\end{align*}
Here, the first inequality comes from the bounds $|\eta_i| \le \frac{4}{\pi}~\forall
i\in\mathbb{N}$ and (\ref{eq:BM:hat_t(n) expression}), and the second inequality comes from the
bounds $i^2 \ge i~\forall i\in\mathbb{N}$ and $n^{\epsilon} \ge (\hat{n})^{\epsilon}~\forall n\ge
\hat{n}$. Note that the series $\sum_{i=1}^{\infty} \hat{M}_i$ converges since it is a geometric
series with a common ratio $\exp(-\hat{m}) \in (0,1)$. Since the target of the functions is a
complete normed vector space, the infinite series $\hat{S}(n)$ converges uniformly on
$\hat{\mathcal{D}}$. Consequently, we can interchange the order of limit and summation, and we
have
\begin{align*}
\lefteqn{\lim_{n \to \infty} \pr\{T_x( c_d \sqrt{n}) \le \hat{t}(n)\}} \\
&= 1 - \lim_{n \to \infty} \hat{S}(n) \\
&= 1 - \sum_{i=1}^{\infty} \eta_i \lim_{n \to \infty} \exp \left( - \frac{\rho_i}{(c_d)^2 n}
\hat{t}(n) \right).
\end{align*}
Since $\hat{t}(n)=\Theta(n^{1+\epsilon})$, we furthermore have
\begin{align*}
\lim_{n \to \infty} \exp \left( - \frac{\rho_i}{(c_d)^2 n} \hat{t}(n) \right)
&= 0,
\end{align*}
which gives $\lim_{n \to \infty} \pr\{T_x( c_d \sqrt{n}) \le \hat{t}(n)\}= 1$. This completes the
proof for $\alpha=2$.

Next, consider the case of $\alpha\in(0,2)$. Similarly to the proof for $\alpha=2$, we can prove
this case by substituting $r= c_d\sqrt{n}$ and $t=\hat{t}(n)$ into (\ref{eq:FPT_distribution_LF})
and showing that
\begin{align}\label{eq:LF:upper}
\lim_{n\to\infty} \pr\{T_x(c_d\sqrt{n}) \le \hat{t}(n)\} = 1.
\end{align}
Since the dominant decay rate $|\lambda_{1}|$ scales as $\Theta(r^{-\alpha}) =
\Theta(n^{-\frac{\alpha}{2}})$, by using approaches in the proof for $\alpha=2$, we can show
(\ref{eq:LF:upper}). Due to similarities, we omit the details. \hfill$ \square$
\end{lemma}

%
%
\smallskip
\begin{lemma}[Lower bound for L\'{e}vy flight]\label{lemma:scaling_property_2_LF} Suppose that under L\'{e}vy flight with a distribution parameter $\alpha\in(0,2]$, the time $t\triangleq \tilde{t}(n)$ in $\pr\{T(c_d \sqrt{n}) > t\}$ scales as $\Theta(n^{\frac{\alpha}{2} - \epsilon})$ for an arbitrary $\epsilon > 0$. Then, we have
\begin{align*}
\lim_{n\to\infty} \pr\{T(c_d\sqrt{n}) > \tilde{t}(n)\} = 1,
\end{align*}
which shows that the critical delay $C_{\Pi}(n)$ under L\'{e}vy flight scales as $\Omega(n^{\frac{\alpha}{2}})$.

\smallskip
\noindent\textit{Proof:} We will prove this lemma by showing that $\lim_{n \to \infty}
\pr\{T_x(c_d\sqrt{n}/\sqrt{2}) \le \tilde{t}(n)\} = 0$. Then, by substituting $r = c_d\sqrt{n}$
and $t= \tilde{t}(n)$ into~(\ref{eq:lower_and_upper_prob}) and taking a limit to $n$, we obtain
\begin{align*}
\lim_{n \to \infty} \pr\{T(c_d\sqrt{n}) \le \tilde{t}(n)\}
&\le \!2\!\lim_{n \to \infty} \pr\{T_x(c_d\sqrt{n}/\sqrt{2}) \le \tilde{t}(n)\} \\
&= 0.
\end{align*}
That is, $\lim_{n \to \infty} \pr\{T(c_d\sqrt{n}) \le \tilde{t}(n)\} = 0$, or equivalently,
$\lim_{n \to \infty} \pr\{T(c_d\sqrt{n}) > \tilde{t}(n)\} = 1$, which proves the lemma.

First, consider the case of $\alpha=2$. We substitute $r = c_d\sqrt{n}/\sqrt{2}$ and $t=
\tilde{t}(n)$ into~(\ref{eq:FPT_distribution_LF}). Then, the series on the right-hand side
of~(\ref{eq:FPT_distribution_LF}) becomes a function of $n$, and analogously to the proof of
Lemma~\ref{lemma:scaling_property_1_LF}, we let
\begin{align*}
\pr\{T_x(c_d\sqrt{n}/\sqrt{2}) \le \tilde{t}(n)\} &= 1-\sum_{i=1}^{\infty} \eta_i \exp \left( -
\frac{2\rho_i}{(c_d)^2 n} \tilde{t}(n) \right) \\
&\deq 1- \tilde{S}(n).
\end{align*}
Similarly to the proof of Lemma~\ref{lemma:scaling_property_1_LF}, we will show that there exists a
constant $\tilde{n}\in\mathbb{N}$ such that the infinite series $\tilde{S}(n)$ converges
uniformly on $\tilde{\mathcal{D}}\deq[\tilde{n},\infty)$.

Since $\tilde{t}(n)=\Theta(n^{1-\epsilon})$, there exist constants $\tilde{n}\in\mathbb{N}$ and
$\tilde{c}>0$ such that
\begin{align}\label{eq:BM:tilde_t(n) expression}
\tilde{t}(n) \ge \tilde{c} n^{1-\epsilon} \quad \text{ for all } n \ge \tilde{n}.
\end{align}
For a technical purpose for showing the uniform convergence, we restrict the domain of $n$ as
$\tilde{\mathcal{D}}_d \triangleq [\tilde{n},\,d]$ for an arbitrary $d \ge \tilde{n}$. Let
$\tilde{m}\deq F(\pi/\sqrt{2}c_d)^2 \tilde{c}d^{-\epsilon}$. Then, the $i$-th function of the
series $\tilde{S}(n)$ is bounded by a constant $\tilde{M}_i\triangleq \frac{4}{\pi}
\{\exp(-\tilde{m})\}^i$ for all $n\in\tilde{\mathcal{D}}_d$ as follows:
\begin{align*}
\bigg|\eta_i \exp \left( - \frac{2\rho_i}{(c_d)^2 n} \tilde{t}(n) \right)\bigg|
&\le \frac{4}{\pi} \exp\left(-\frac{2\rho_i}{(c_d)^2n} \tilde{c} n^{1-\epsilon}\right) \\
&\le \frac{4}{\pi} \exp\left(-\frac{ Fi (\pi)^2 }{2(c_d)^2} \tilde{c} d^{-\epsilon}\right) \\
&= \tilde{M}_i.
\end{align*}
Here, the first inequality comes from the bounds $|\eta_i| \le \frac{4}{\pi}~\forall
i\in\mathbb{N}$ and (\ref{eq:BM:tilde_t(n) expression}), and the second inequality comes from the
bounds $i^2 \ge i~\forall i\in\mathbb{N}$ and $n^{-\epsilon} \ge d^{-\epsilon}~\forall n
\in\tilde{\mathcal{D}}_d$. Note that the series $\sum_{i=1}^{\infty} \tilde{M}_i$ converges since
it is a geometric series with a common ratio $\exp(-\tilde{m}) \in (0,1)$. Hence, the infinite
series $\tilde{S}(n)$ converges uniformly on $\tilde{\mathcal{D}}_d$. Since $d$ is arbitrary, we
get uniform convergence on~$\tilde{\mathcal{D}}$. Consequently, we can interchange the order of
limit and summation, and we have
\begin{align*}
\lefteqn{\lim_{n \to \infty} \pr\{T_{x}(c_d \sqrt{n}/\sqrt{2}) \le \tilde{t}(n)\}} \\
&= 1 - \lim_{n \to \infty} \tilde{S}(n) \\
&= 1- \sum_{i=1}^{\infty} \eta_i \lim_{n\to\infty} \exp \left( - \frac{2\rho_i}{(c_d)^2
n}\tilde{t}(n)\right).
\end{align*}
Since $\tilde{t}(n)=\Theta(n^{1-\epsilon})$, we furthermore have
\begin{align*}
\lim_{n\to\infty} \exp \left( - \frac{2\rho_i}{(c_d)^2 n}\tilde{t}(n)\right)
&=1,
\end{align*}
which gives
\begin{align*}
\lim_{n \to \infty} \pr\{T_{x}(c_d \sqrt{n}/\sqrt{2}) \le \tilde{t}(n)\} = 1 -
\sum_{i=1}^{\infty} \eta_i.
\end{align*}
Note from~(\ref{eq:FPT_distribution_LF}) that $\pr\{T_x(c_d \sqrt{n}/\sqrt{2}) \le 0\} =  1 -
\sum_{i=1}^{\infty} \eta_i$. In addition, it is obvious that $\pr\{T_{x}(c_d\sqrt{n}/\sqrt{2})
\le 0\} = 0$. Therefore, we have $\lim_{n \to \infty} \pr\{T_{x}(c_d\sqrt{n}/\sqrt{2}) \le
\tilde{t}(n)\} = 0$. This completes the proof for $\alpha=2$.

Next, consider the case of $\alpha\in(0,2)$. Similarly to the proof for $\alpha=2$, we can prove
this case by substituting $r= c_d\sqrt{n}/\sqrt{2}$ and $t=\tilde{t}(n)$ into
(\ref{eq:FPT_distribution_LF}) and showing that
\begin{align}\label{eq:LF:lower}
\lim_{n\to\infty} \pr\{T_x(c_d\sqrt{n}/\sqrt{2}) \le \tilde{t}(n)\} = 0.
\end{align}
Since the dominant decay rate $|\lambda_{1}|$ scales as $\Theta(r^{-\alpha}) =
\Theta(n^{-\frac{\alpha}{2}})$, by using approaches in the proof for $\alpha=2$, we can show
(\ref{eq:LF:lower}). Due to similarities, we omit the details. \hfill $\square$
\end{lemma}
\smallskip

Combining Lemmas~\ref{lemma:scaling_property_1_LF} and~\ref{lemma:scaling_property_2_LF} yields
the following theorem.

\smallskip
\begin{theorem}\label{thm:critical_delay_LF}
\textbf{The critical delay} $\bm{C_{\Pi}\text{\bf{(}}n\text{\bf{)}}}$ \textbf{under L\'{e}vy
flight with a distribution parameter} \text{\boldmath$\boldmath{\alpha\in(0,2]}$\unboldmath}
\textbf{scales as} $\bm{\Theta
\text{\bf{(}}n^{\frac{\text{\boldmath$\alpha$\unboldmath}}{2}}\text{\bf{)}}}$.
\end{theorem}
\smallskip

\begin{remark}
The main idea behind the proof of Lemmas~\ref{lemma:scaling_property_1_LF}
and~\ref{lemma:scaling_property_2_LF} was that the smallest decay rate in the exponential
functions in~(\ref{eq:FPT_distribution_LF}) (i.e., $\frac{\rho_1}{r^\alpha}$) determines the
limiting behavior of the first exit time distribution. That is, the smallest decay rate
characterizes the critical delay under L\'{e}vy flight.
\end{remark}

\section{Critical Delay Analysis for L\'{e}vy Walk}\label{sec:analysis:LW}
In this section, we will show that the critical delay $C_{\Pi}(n)$ under L\'{e}vy walk with a
distribution parameter $\alpha$ scales as $\Theta(n^{\frac{1}{2}})$ for $\alpha\in(0,1)$ and
$\Theta(n^{\frac{\alpha}{2}})$ for $\alpha\in[1,2]$ (Theorem~\ref{thm:critical_delay_LW}). In
Section~\ref{sec:analysis:LW:tech}, we explain technical challenges and our approach for proving
Theorem~\ref{thm:critical_delay_LW}. In Section~\ref{sec:analysis:LW:ana}, we prove
Theorem~\ref{thm:critical_delay_LW} by showing that the upper bound on $C_{\Pi}(n)$ scales as
$O(n^{\frac{1}{2}})$ for $\alpha\in(0,1)$ and $O(n^{\frac{\alpha}{2}})$ for $\alpha\in[1,2]$
(Lemma~\ref{lemma:scaling_property_1_LW}), and that the lower bound on $C_{\Pi}(n)$ scales as
$\Omega(n^{\frac{1}{2}})$ for $\alpha\in(0,1)$ and $\Omega(n^{\frac{\alpha}{2}})$ for
$\alpha\in[1,2]$ (Lemma~\ref{lemma:scaling_property_2_LW}).

\subsection{Technical Approach}\label{sec:analysis:LW:tech}
We first explain the technical challenges that preclude the use of our technique for L\'{e}vy
flight as well as other conventional techniques. We next explain our technical approach to deal
with these challenges. The technical challenges are two-folds and are mainly inherent in the
L\'{e}vy walk nature.

(i) We begin with the description of differences between L\'{e}vy flight and L\'{e}vy walk from a
modeling perspective. Let $t_i\,(i=1,2,\ldots)$ denote the time instant when the $i$-th step
begins. We take the time $t_i$ as the embedded point of the process $\{\bm{X}(t)\}_{t \ge 0}$,
and focus on the corresponding embedded process $\{\bm{E}_i\}_{i\in\mathbb{N}} \deq
\{\bm{X}(t_i)\}_{i\in\mathbb{N}}$. Under both L\'{e}vy mobility models, at each embedded point
$t_i$, the destination of the next step of the $i$-th step (i.e., $\bm{E}_{i+1}$) is chosen
independently of the past locations at time $t < t_{i}$ and depends only on the current location
at time $t = t_i$. That is, the embedded process $\{\bm{E}_i\}_{i\in\mathbb{N}}$ satisfies the
following Markov property:
\begin{align*}
\pr\{ &\bm{E}_{i+1} = \bm{x}_{i+1}\,|\,\bm{E}_{j} = \bm{x}_j, j = 1,\ldots,i \} \\
&=\pr\{\bm{E}_{i+1} = \bm{x}_{i+1}\,|\,\bm{E}_{i} = \bm{x}_i\}.
\end{align*}
Thus, under both L\'{e}vy mobility models, the process $\{\bm{X}(t_i)\}_{i\in\mathbb{N}}$ becomes
a discrete-time Markov chain. However, the fact that the embedded point $t_i$ is chosen in a
different way for L\'{e}vy flight and L\'{e}vy walk incurs the key challenge. In the case of
L\'{e}vy flight, it is chosen deterministically as $t_i = i-1$. Therefore, L\'{e}vy flight is a
discrete-time Markov process. However, in the case of the L\'{e}vy walk, the embedded point is
chosen stochastically and is correlated with step size as follows: $t_i = \sum_{j=1}^{i-1} Z_j$
(where $Z_j$ is a random variable denoting the $j$-th step size). Therefore, the L\'{e}vy walk is
a semi-Markov process~\cite{RW01Drysdale} whose embedded process becomes L\'{e}vy flight.

(ii) The proof of Lemma~\ref{lemma:1D_projection_LF} also shows that, for a given 2-dimensional
L\'{e}vy walk, its 1-dimensional projected processes also have a power-law type of step size
distribution. However, the velocity of the projected processes is not a constant for every step,
which implies that neither of 1-dimensional projected processes of 2-dimensional L\'{e}vy walk
can be 1-dimensional L\'{e}vy walk.

Consequently, the technique used for L\'{e}vy flight in this paper is not applicable because it
requires decoupling of space and time. In addition, the occupation probability $P(x,t)$ is not
available and the derivation is not mathematically tractable.

To cope with these technical challenges, we propose a different approach based on a stochastic
analysis technique characterizing the embedded Markov process of a semi-Markov process.
Specifically, our approach is to derive a relation between the first exit time under L\'{e}vy
flight (i.e., embedded Markov process) and that under L\'{e}vy walk (i.e., semi-Markov process).
From this relation, our technique derives a tight upper bound for the critical delay. Then, by
combining the upper bound and a lower bound for the critical delay inferred from our analytical
result of L\'{e}vy flight in Section~\ref{sec:analysis:LF}, we can provide the exact order of the critical delay under L\'{e}vy
walk.

\subsection{Analysis}\label{sec:analysis:LW:ana}
\begin{figure}[t!]
  \centering
  {\epsfig{figure=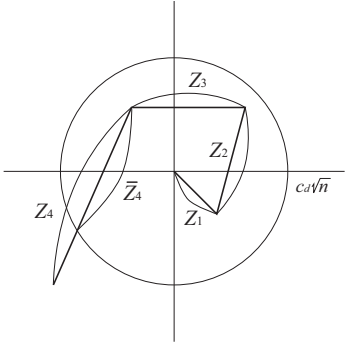,width=0.25\textwidth,height=0.25\textwidth}}
   \caption{An example of the random variables $N(n)\,(=4)$, $Z_i$ and $\bar{Z}_{N(n)}$.}
   \label{fig:proof}
\end{figure}

Let $N(n)$ be a random variable denoting the number of steps occurred until $t \le
T(c_d\sqrt{n})$. Then,
\begin{align}\label{eq:LW:fet}
\begin{split} T(c_d\sqrt{n}) =
\begin{cases} c_d\sqrt{n} &\text{ if } N(n) = 1, \\
\sum_{i=1}^{N(n)-1} Z_i + \bar{Z}_{N(n)} & \text{ if } N(n) \ge 2,
\end{cases}
\end{split}
\end{align}
where $\bar{Z}_{N(n)}$ is a random variable denoting the moving distance during the $N(n)$-th
step until exiting the disc $\mathcal{D}(n)$ (See Fig.~\ref{fig:proof}.). Note that
$\bar{Z}_{N(n)}$ is not identically distributed with $Z_i$, and we have
\begin{align}\label{eq:LW:bound wpb 1}
\bar{Z}_{N(n)} < 2c_d\sqrt{n}~~\text{ with probability 1}.
\end{align}
The random variable $N(n)$ is closely related to the first exit time under L\'{e}vy flight,
denoted by $T_{\text{LF}}(c_d\sqrt{n})$, as follows:
\begin{align}\label{eq:LW:equal distribution}
N(n) \ed \lceil T_{\text{LF}}(c_d\sqrt{n}) \rceil,
\end{align}
where $\lceil x \rceil$ denotes the smallest integer larger than or equal to~$x$. In
Lemma~\ref{lemma:scaling_property_N}, we derive the order of $\E[N(n)]$, which will be used to
study the critical delay under L\'{e}vy walk.

\smallskip
\begin{lemma}\label{lemma:scaling_property_N} $\E[N(n)]$ scales as $\Theta(n^{\frac{\alpha}{2}})$
for $\alpha\in(0,2]$.

\smallskip
\noindent\textit{Proof:} From Lemma~\ref{lemma:scaling_property_1_LF}, we have $\lim_{n\to\infty}\pr\{T_{\text{LF}}(c_d\sqrt{n})\le \hat{t}(n) \} = 1$ when $\hat{t}(n) = \Theta(n^{\frac{\alpha}{2}+\hat\epsilon})$ for $\alpha\in(0,2]$ and an arbitrary $\hat\epsilon >0$. Hence, we have
\begin{align}\label{eq:LW:E_N_upper}
\E[T_{\text{LF}}(c_d\sqrt{n})] = O(n^{\frac{\alpha}{2}+\hat\epsilon}).
\end{align}
From Lemma~\ref{lemma:scaling_property_2_LF}, we have $\lim_{n\to\infty}\pr\{T_{\text{LF}}(c_d\sqrt{n}) > \tilde{t}(n)\} = 1$ when $\tilde{t}(n) =
\Theta(n^{\frac{\alpha}{2}-\tilde\epsilon})$ for $\alpha\in(0,2]$ and an arbitrary $\tilde\epsilon>0$. Thus, we have
\begin{align}\label{eq:LW:E_N_lower}
\E[T_{\text{LF}}(c_d\sqrt{n})] = \Omega(n^{\frac{\alpha}{2}-\tilde\epsilon}).
\end{align}
By choosing $\hat{\epsilon}$ and $\tilde{\epsilon}$ arbitrarily small, from
(\ref{eq:LW:E_N_upper}) and (\ref{eq:LW:E_N_lower}), we have
\begin{align}\label{eq:LW:order_E_N_1}
\E[T_{\text{LF}}(c_d\sqrt{n})] = \Theta(n^{\frac{\alpha}{2}}) \quad \forall \alpha \in (0,2].
\end{align}
Note from (\ref{eq:LW:equal distribution}) that
\begin{align}\label{eq:LW:order_E_N_2}
\E[T_{\text{LF}}(c_d\sqrt{n})] \le \E[N(n)] \le \E[T_{\text{LF}}(c_d\sqrt{n})]+1,
\end{align}
which shows that the order of $\E[N(n)]$ is the same as that of $\E[T_{\text{LF}}(c_d\sqrt{n})]$.
Therefore, combining (\ref{eq:LW:order_E_N_1}) and (\ref{eq:LW:order_E_N_2}) yields the lemma.
\hfill $\square$
\end{lemma}
\smallskip

With the help of Lemma~\ref{lemma:scaling_property_N}, we can derive an upper bound for the
critical delay under L\'{e}vy walk.

\smallskip
\begin{lemma}[Upper bound for L\'{e}vy walk] \label{lemma:scaling_property_1_LW} Suppose that under L\'{e}vy walk with a distribution parameter $\alpha$, the time $t\triangleq \hat{t}(n)$ in $\pr\{T(c_d \sqrt{n}) > t\}$ scales as $\Theta(n^{\frac{1}{2} + \epsilon_1})$ for an arbitrary $\epsilon_1 > 0$ and $\alpha \in (0,1)$, and $\Theta(n^{\frac{\alpha}{2} + \epsilon_2})$ for an arbitrary $\epsilon_2 > 0$ and $\alpha \in [1,2]$. Then, we have
\begin{align*}
\lim_{n\to\infty} \pr\{T(c_d\sqrt{n}) >
\hat{t}(n)\} = 0,
\end{align*}
which shows that the critical delay $C_{\Pi}(n)$ under L\'{e}vy walk scales as $O(n^{\frac{1}{2}})$ for $\alpha \in (0,1)$ and $O(n^{\frac{\alpha}{2}})$ for $\alpha\in [1,2]$.

\smallskip
\noindent\textit{Proof:} Using Markov's inequality~\cite{Ross1996}, we have
\begin{align}\label{eq:LW:markov}
\pr\{T(c_d\sqrt{n}) > \hat{t}(n)\} \le \frac{\E[T(c_d\sqrt{n})]}{\hat{t}(n)}.
\end{align}
We calculate the expectation $\E[T(c_d\sqrt{n})]$ on the right-hand side of (\ref{eq:LW:markov})
by conditioning on the values of $N(n)$ as
\begin{align}\label{eq:LW:exp_T}
\E[T(c_d\sqrt{n})]
&= \E\!\left[\E[T(c_d\sqrt{n})\,|\,N(n)]\right] \nonumber \\
&= \sum_{k=1}^{\infty}\E[T(c_d\sqrt{n})\,|\,N(n)=k] \cdot \pr\{N(n)=k\}.
\end{align}
From~(\ref{eq:LW:fet}), we have for $k=1$,
\begin{align}\label{eq:LW:conditional exp_k_1}
\E[T(c_d\sqrt{n})\,|\,N(n)=k] &= c_d\sqrt{n}.
\end{align}
In addition, from (\ref{eq:LW:fet}), we have for $k=2,3,\ldots$,
\begin{align}\label{eq:LW:conditional exp_k_2}
\lefteqn{\E[T(c_d\sqrt{n})\,|\,N(n)=k]} \nonumber \\
&= \E\left[\sum_{i=1}^{N(n)-1} Z_i + \bar{Z}_{N(n)}
\,\bigg|\,N(n)=k\right] \nonumber \\
&= \sum_{i=1}^{k-1} \E[Z_i \,|\,N(n)=k] + \E[\bar{Z}_k \,|\,N(n)=k].
\end{align}
The random variables $Z_i\,(i=1,\ldots,k-1)$ and $\bar{Z}_k$ in (\ref{eq:LW:conditional exp_k_2})
are correlated with the random variable $N(n)\,(=k)$, whereas the random variables
$Z_i\,(i=k+1,k+2,\ldots)$ are independent of $N(n)\,(=k)$. Specifically, for $i=1,\ldots, k-1$,
the step size $Z_i$ should be less than the diameter of the disc $\mathcal{D}(n)$ (i.e.,
$2c_d\sqrt{n}$). In addition, the truncated step size $\bar{Z}_k$ should satisfy the inequality
in (\ref{eq:LW:bound wpb 1}). Hence, the conditional expectations on the right-hand side of
(\ref{eq:LW:conditional exp_k_2}) are bounded as follows:
\begin{align}\label{eq:LW:conditional_exp_k_2_2}
\begin{split}
&\E[Z_i | N(n)=k] \le \E[Z | Z\le 2c_d\sqrt{n}], \\
&\E[\bar{Z}_k | N(n)=k] \le 2 c_d \sqrt{n},
\end{split}
\end{align}
where $Z$ denotes the generic random variable for $Z_i$.\footnote{$\E[Z_i \,|\, N(n)=k] = \E[Z]
\text{ for } i = k+1,k+2,\cdots.$} Combining
(\ref{eq:LW:exp_T})-(\ref{eq:LW:conditional_exp_k_2_2}), we obtain an upper bound for
$\E[T(c_d\sqrt{n})]$ as follows:
\begin{align}\label{eq:LW:exp_T_bound}
\lefteqn{\E[T(c_d\sqrt{n})]} \nonumber \\
&\le c_d\sqrt{n} \cdot \pr\{N(n)=1\} + 2c_d \sqrt{n} \sum_{k=2}^{\infty}\pr\{N(n)=k\} \nonumber
\\
& \quad + \E[Z|Z \le 2c_d\sqrt{n}] \sum_{k=2}^{\infty} (k-1) \cdot \pr\{N(n)=k\} \nonumber \\
& \le 2c_d \sqrt{n} \sum_{k=1}^{\infty}\pr\{N(n)=k\} \nonumber \\
& \quad + \E[Z |Z \le 2 c_d\sqrt{n}]
\sum_{k=1}^{\infty} k \cdot \pr\{N(n)=k\} \nonumber \\
&=2c_d \sqrt{n} + \E[Z |Z \le 2 c_d\sqrt{n}] \cdot \E[N(n)].
\end{align}
Using (\ref{eq:model:ccdf}), we can calculate the conditional expectation $\E[Z|Z \le 2
c_d\sqrt{n}]$ in (\ref{eq:LW:exp_T_bound}) and it scales for each $\alpha\in(0,2]$ as follows:
\begin{align*}
\lefteqn{\E[Z|Z \le 2 c_d\sqrt{n}]} \\
&=
\begin{cases}
\frac{\alpha}{1-\alpha} \frac{(2c_d\sqrt{n})^{1-\alpha}-1}{1-(2c_d\sqrt{n})^{-\alpha}} & \text{
for } \alpha \in (0, 1), \\
\frac{\log(2c_d\sqrt{n})}{1-(2c_d\sqrt{n})^{-1}} & \text{ for } \alpha = 1, \\
\frac{\alpha}{\alpha-1} \frac{1-(2c_d\sqrt{n})^{1-\alpha}}{1-(2c_d\sqrt{n})^{-\alpha}} & \text{
for } \alpha \in (1, 2), \\
\frac{\sqrt{2} \sigma}{\sqrt{\pi}} \frac{\exp(-1/2\sigma^2)-\exp(-2(c_d)^2
n/\sigma^2)}{\mathrm{erf}(c_d\sqrt{2n}/\sigma)-\mathrm{erf}(1/\sqrt{2}\sigma)} & \text{ for }
\alpha = 2, \\
\end{cases} \\
&=
\begin{cases}
\Theta(n^{(1-\alpha)/2}) & \text{ for } \alpha \in (0, 1), \\
\Theta(\log(n)) & \text{ for } \alpha = 1, \\
\Theta(n^{0}) & \text{ for } \alpha \in (1, 2].
\end{cases}
\end{align*}
Since $\E[N(n)]$ scales as $\Theta(n^{\frac{\alpha}{2}})$ by
Lemma~\ref{lemma:scaling_property_N}, the term on the right-hand side of
(\ref{eq:LW:exp_T_bound}) scales as
\begin{align}\label{eq:LW:order}
\lefteqn{2c_d \sqrt{n} + \E[Z |Z \le 2 c_d\sqrt{n}] \cdot \E[N(n)]} \nonumber \\
& = \begin{cases}
\Theta(n^{\frac{1}{2}}) & \text{ for } \alpha \in (0,1), \\
\Theta(n^{\frac{1}{2}}\log(n)) & \text{ for } \alpha =1, \\
\Theta(n^{\frac{\alpha}{2}}) & \text{ for } \alpha \in (1,2].
\end{cases}
\end{align}
Thus, we have from (\ref{eq:LW:exp_T_bound}) and (\ref{eq:LW:order}) the following:
\begin{align*}
\lefteqn{\lim_{n\to\infty} \frac{\E[T(c_d\sqrt{n})]}{\hat{t}(n)}}\\
&\le \lim_{n\to\infty} \frac{2c_d \sqrt{n} + \E[Z |Z \le 2 c_d\sqrt{n}]  \cdot
\E[N(n)]}{\hat{t}(n)}\\
& = 0.
\end{align*}
Therefore, from (\ref{eq:LW:markov}), we have
\begin{align*}
\lim_{n\to\infty} \pr\{T(c_d\sqrt{n}) >
\hat{t}(n)\} \le \lim_{n\to\infty} \frac{\E[T(c_d\sqrt{n})]}{\hat{t}(n)}  \le 0,
\end{align*}
i.e., $\lim_{n\to\infty} \pr\{T(c_d\sqrt{n}) > \hat{t}(n)\}= 0$. This completes the proof. \hfill
$\square$
\end{lemma}

\smallskip
\begin{lemma}[Lower bound for L\'{e}vy walk]\label{lemma:scaling_property_2_LW} Suppose that under L\'{e}vy walk with a distribution parameter $\alpha$, the time $t\triangleq \tilde{t}(n)$ in $\pr\{T(c_d \sqrt{n}) > t\}$ scales as $\Theta(n^{\frac{1}{2} - \epsilon_1})$ for an arbitrary $\epsilon_1 > 0$ and $\alpha\in(0,1)$, and $\Theta(n^{\frac{\alpha}{2} - \epsilon_2})$ for an arbitrary $\epsilon_2 > 0$ and $\alpha\in(1,2]$. Then, we have
\begin{align*}
\lim_{n\to\infty} \pr\{T(c_d\sqrt{n}) >
\tilde{t}(n)\} = 1,
\end{align*}
which shows that the critical delay $C_{\Pi}(n)$ under L\'{e}vy walk scales as $\Omega(n^{\frac{1}{2}})$ for $\alpha \in (0,1)$ and $\Omega(n^{\frac{\alpha}{2}})$ for $\alpha\in [1,2]$.

\smallskip
\noindent\textit{Proof:} We will prove this lemma by showing for each of the cases of
$\alpha\in(0,1)$ and $\alpha\in[1,2]$ that
\begin{align*}
\lim_{n\to\infty}\pr\{T(c_d\sqrt{n}) \le \tilde{t}(n)\} = 0.
\end{align*}
We first consider the case of $\alpha \in (0,1)$. Since a L\'{e}vy walker moves with a constant
velocity $v=1$, it takes at least $c_d\sqrt{n}$ time to exit from the disc $\mathcal{D}(n)$.
Thus, it is obvious that
\begin{align}\label{eq:LW:lower:bd1}
\pr \{T(c_d\sqrt{n}) < c_d\sqrt{n}\} = 0.
\end{align}
Since $\tilde{t}(n) = \Theta(n^{\frac{1}{2}-\epsilon_1})$, there exists a constant
$\tilde{n}\in\mathbb{N}$ such that $\tilde{t}(n) < c_d \sqrt{n}$ for $n \ge \tilde{n}$. Hence, we
have for $n \ge \tilde{n}$
\begin{align}\label{eq:LW:lower:bd2}
\pr\{T(c_d\sqrt{n}) \le \tilde{t}(n)\} &\le
\pr \{T(c_d\sqrt{n}) < c_d\sqrt{n}\}.
\end{align}
Combining (\ref{eq:LW:lower:bd1}) and (\ref{eq:LW:lower:bd2}) and then taking limits, we have
\begin{align*}
\lim_{n\to\infty} \pr\{T(c_d\sqrt{n}) \le \tilde{t}(n)\} &\le
\lim_{n\to\infty} \pr \{T(c_d\sqrt{n}) < c_d\sqrt{n}\} \\
&= 0,
\end{align*}
i.e., $\lim_{n\to\infty}\pr\{T(c_d\sqrt{n}) \le \tilde{t}(n)\} = 0$. We have proved the lemma in
the case of $\alpha \in (0,1)$.

We next consider the case of $\alpha \in [1,2]$. In the following, we use the notations
$T_{\text{LF}}(\cdot)$ and $T_{\text{LW}}(\cdot)$ to distinguish the first exit times between
L\'{e}vy flight and L\'{e}vy walk. We will show based on (\ref{eq:LW:fet}) and (\ref{eq:LW:equal
distribution}) that for $t \ge 0$,
\begin{align}\label{eq:LW:lower_prob_bound}
\pr\{T_{\text{LW}}(c_d\sqrt{n}) \le t\} \le \pr\{T_{\text{LF}}(c_d\sqrt{n}) \le t+1\}.
\end{align}
From (\ref{eq:LW:fet}), if $N(n)=1$, then $T_{\text{LW}}(c_d\sqrt{n}) = c_d\sqrt{n} > 0 =
N(n)-1.$ In addition, if $N(n) \ge 2$, then $T_{\text{LW}}(c_d\sqrt{n}) = \sum_{i=1}^{N(n)-1} Z_i
+ \hat{Z}_N > \sum_{i=1}^{N(n)-1} Z_i \ge N(n)-1,$ where the last inequality comes from the
assumption that the step size~$Z$ has a lower bound at 1 (given in
Section~\ref{sec:levymobility}). Combining above two cases gives
\begin{align*}
T_{\text{LW}}(c_d\sqrt{n}) \ge N(n) -1.
\end{align*}
From (\ref{eq:LW:equal distribution}), we obtain $N(n)\ed \lceil T_{\text{LF}}(c_d\sqrt{n})
\rceil \ge T_{\text{LF}}(c_d\sqrt{n})$. Thus, we have with probability 1,
\begin{align*}
T_{\text{LW}}(c_d\sqrt{n}) \ge T_{\text{LF}}(c_d\sqrt{n}) -1.
\end{align*}
This proves (\ref{eq:LW:lower_prob_bound}). Substituting $t=\tilde{t}(n)$ into
(\ref{eq:LW:lower_prob_bound}), we obtain
\begin{align}\label{eq:LW:lower_prob_bound_sub}
\pr\{T_{\text{LW}}(c_d\sqrt{n}) \le \tilde{t}(n)\} \le \pr\{T_{\text{LF}}(c_d\sqrt{n}) \le
\tilde{t}(n)\!+\!1\}.
\end{align}
For $\tilde{t}(n)$ scaling as $\Theta(n^{\frac{\alpha}{2}-\epsilon_2})$, $\tilde{t}(n)+1$ also
scales as $\Theta(n^{\frac{\alpha}{2}-\epsilon_2})$. Consequently, by
Lemma~\ref{lemma:scaling_property_2_LF}, the probability on the right-hand side of
(\ref{eq:LW:lower_prob_bound_sub}) becomes in the limit:
\begin{align*}
\lim_{n\to\infty}\pr\{T_{\text{LF}}(c_d\sqrt{n}) \le \tilde{t}(n)+1\} = 0.
\end{align*}
Therefore, from (\ref{eq:LW:lower_prob_bound_sub}), we have
$\lim_{n\to\infty}\pr\{T_{\text{LW}}(c_d\sqrt{n}) \le \tilde{t}(n)\} = 0$, which proves the lemma
in the case of $\alpha \in [1,2]$. This completes the proof. \hfill$ \square$
\end{lemma}

\smallskip
Combining Lemmas~\ref{lemma:scaling_property_1_LW} and~\ref{lemma:scaling_property_2_LW} yields
the following theorem.

\smallskip
\begin{theorem}\label{thm:critical_delay_LW} {\bf The critical delay}
$\bm{C_{\Pi}\text{\bf{(}}n\text{\bf{)}}}$ {\bf under L\'{e}vy walk with a distribution parameter}
\text{\boldmath$\alpha$\unboldmath} {\bf scales as}
$\bm{\Theta\text{\bf(}n^{\frac{1}{2}}\text{\bf)}}$ {\bf for}
\text{\boldmath$\alpha\in(0,1)$\unboldmath} {\bf and}
$\bm{\Theta\text{\bf{(}}n^{\frac{\text{\boldmath{$\alpha$}\unboldmath}}{2}}\text{\bf)}}$ {\bf
for} \text{\boldmath$\alpha\in[1,2].$\unboldmath}
\end{theorem}

\section{Discussion}\label{sec:discussion}
\begin{figure}[t!]
 	\centering
	\subfigure[L\'{e}vy flight]{\epsfig{figure=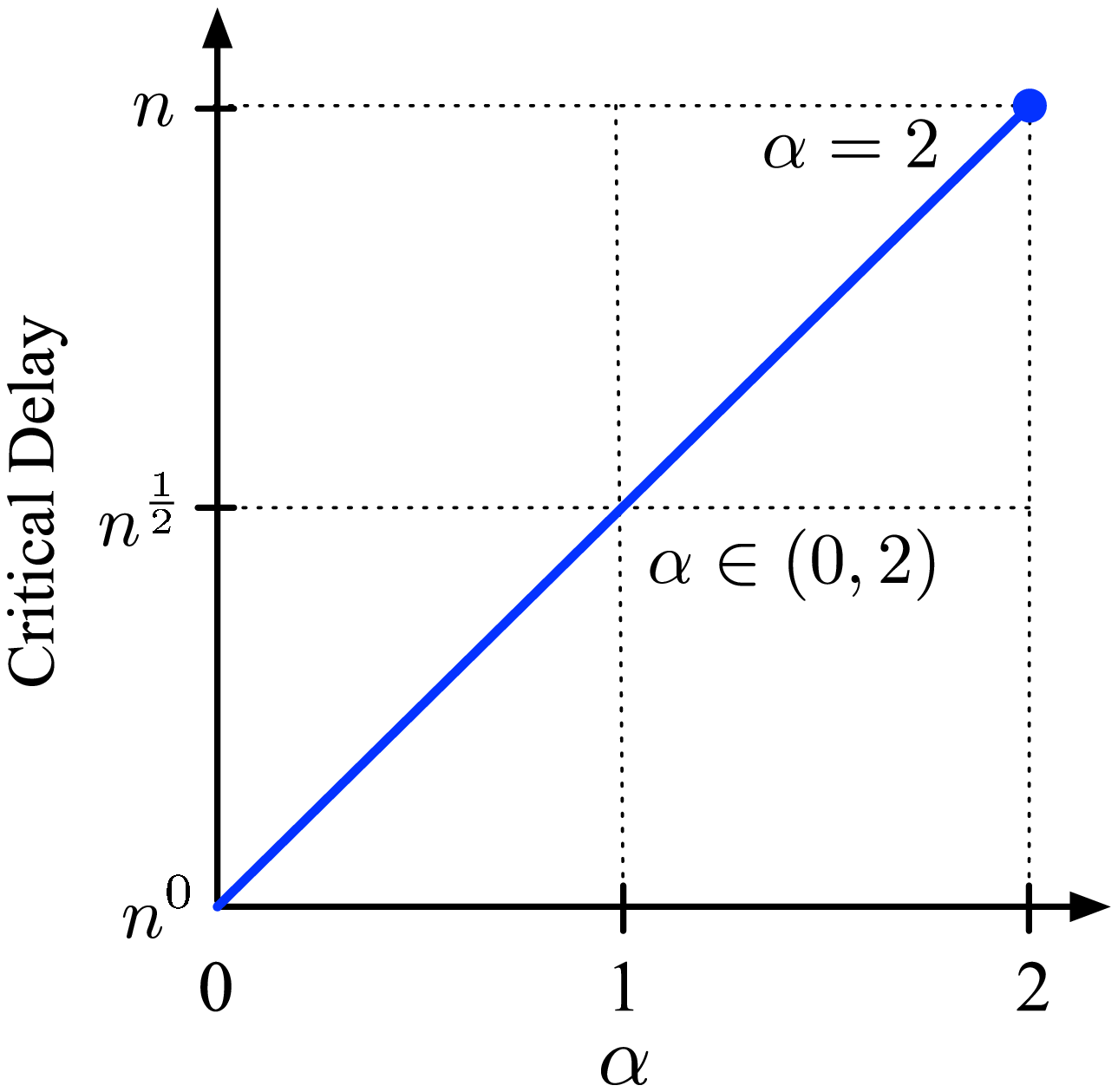,width=0.46\columnwidth}}
	\subfigure[L\'{e}vy walk]{\epsfig{figure=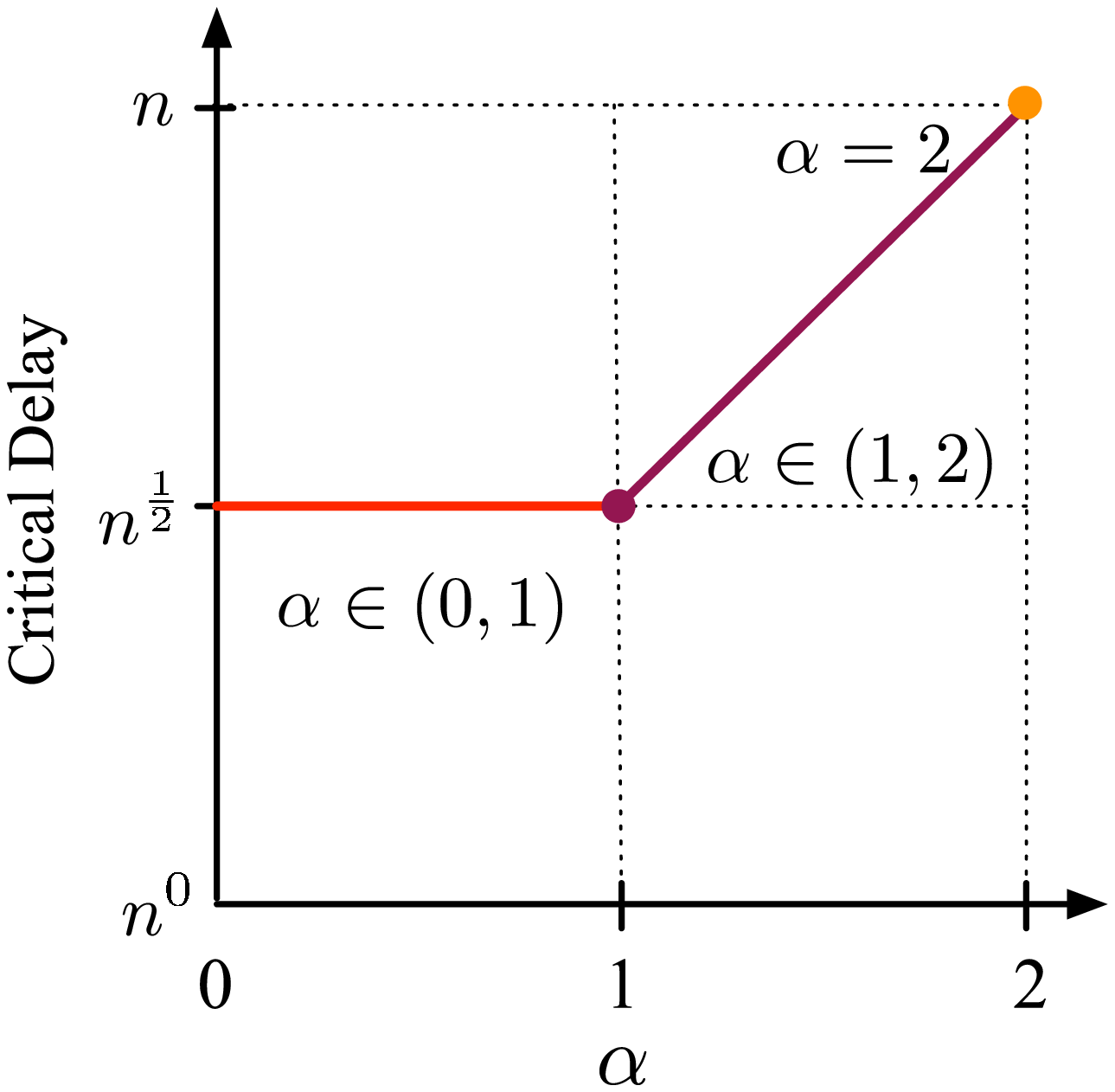,width=0.46\columnwidth}}
	\caption{Critical delays under L\'{e}vy flight and L\'{e}vy walk for different $\alpha.$}
	\label{fig:criticaldelay}
\end{figure}

We summarize the high-level interpretations of this paper. Fig. \ref{fig:criticaldelay}. shows the critical delays under L\'{e}vy walk and L\'{e}vy flight, parameterized by $\alpha.$ L\'{e}vy flight shows that the critical delay proportionally increases with $\alpha.$ However, in the case of the L\'{e}vy walk, we can find a phase transition such that when $\alpha\in(0,1),$ the critical delay is constantly $\Theta (n^{\frac{1}{2}})$ and shifts to the proportional increasing phase when $\alpha \in [1,2].$ Two different scaling regions are essentially related to the fact that the mean step size of L\'{e}vy walk for $\alpha \in (0,1)$ is infinite but finite for $\alpha \in [1,2].$ In contrast to L\'{e}vy walk, the travel time independence of step size in L\'{e}vy flight leads to continuous scaling over $\alpha$. Note that for $\alpha = 2$ (i.e., Brownian motion) our result coincides with that in~\cite{SL_Sharma07} which also studied the critical delay under Brownian motion.

By using values of $\alpha$ from experimental measurements from \cite{rhee:levymobility}, we can see how network delay scales with human mobility in practice. To give an insight to the readers, we show $\alpha$ values measured from five different sites in Table~\ref{tbl:realalpha} presented in~\cite{rhee:levymobility} with a flight extraction method, ``rectangle''.\footnote{We do not present $\alpha$ values from other extraction methods in~\cite{rhee:levymobility} which intentionally exclude some detailed motions of real traces. To capture specific behaviors of humans, one can borrow those $\alpha$ values.} We see that critical delays for human mobility range from $\Theta(n^{0.27})$ to $\Theta (n^{0.91})$. Human mobility mainly has $\alpha > 1$, in which case a longer delay than $\Theta(\sqrt{n})$ is needed. This implies that it may be hard to design a low delay protocol for mobile networks under human mobility.

\begin{table}[t!]
	\caption{Experimental $\alpha$ values for different sites presented in~\cite{rhee:levymobility}.} 
	\centering
	\begin{tabular}{|c|c|c|c|}
	\hline
	Site 		& $\alpha$ 	& Site 		& $\alpha$ \\
	\hline\hline
	KAIST 		& 0.53  	& New York City 	& 1.62 \\
	NCSU 		& 1.27  	& Disney World 		& 1.20 \\
				& 			& State fair 	& 1.81 \\
	\hline
	\end{tabular}
	\label{tbl:realalpha}
\end{table}

Our contribution is not restricted to the mathematical derivation of delay scaling for new mobility models. We provided techniques that connect the diffusion equation of a continuous time random walk to the delay scaling as well as that analyze the delay scaling of semi-Markovian movements. We expect that our techniques can be further developed to the analysis of other detailed performance metrics such as contact time distribution and the generalized delay-capacity tradeoff for various levels of per-node throughput.

Future work includes investigation of throughput and delay scaling for mobile networks with heterogeneous and collective node mobilities. In addition to the recent research topics on ``per-node throughput scaling'' under inhomogeneous spatial node distributions (i.e., Cox process, Neyman-Scott process, Mat\'{e}rn cluster process and Thomas process), e.g., \cite{SL33Garetto, SL35Alfano_j}, our paper can be an important step to the study of delay scaling under such heterogeneous networks. There is an insight from \cite{lee:slaw09} that in human-assisted  networks, the actual delays might be even shorter. This is because human mobility is not completely random: people tend to visit the same locations and regularly meet a group of people every day. Although their mobility can be characterized by heavy-tail distributions, these regularity in daily mobility significantly facilitates routing of packets among people (as long as they are socially connected). Therefore, there remains a possibility of designing a low delay protocol for mobile networks under heterogeneous human mobility by judiciously utilizing these social factors. 
\section{Conclusion} \label{sec:conclusion}
We have presented L\'{e}vy mobility models consisting of L\'{e}vy flight and L\'{e}vy walk parameterized by $\alpha$ and studied the critical delay under both mobility models. L\'{e}vy mobility is known as a realistic human mobility so that the critical delay we provided here can be essential in designing an architecture and protocols of a wireless mobile network. The insight that the critical delay scales as $\Theta (n^{\frac{\alpha}{2}})$ for L\'{e}vy mobility models in the range of $ \alpha \in [1,2]$ is especially important because it is anticipating that the delay of mobile networks with human mobility (e.g., smartphone networks, pocket switched networks) could be quite high in practice, considering the $\alpha$ values measured in real traces. The insight tells that mobile networks operated by human mobility patterns may need to prepare an alternative path for delay sensitive data as well as even for delay tolerable data whose tolerance level is limited. 
\appendices
\section{Critical Delay Analysis for L\'{e}vy Flight without Continuous
Limit}\label{sec:appendix:A}
In Section~\ref{sec:analysis:LF}, we have studied the critical delay under L\'{e}vy flight using
continuous limit. By following the technique in~\cite{SL_Sharma07}, we can study the critical
delay without continuous limit (i.e., with a non-zero scale factor $|c|$) and can derive a
lower bound for the critical delay under L\'{e}vy flight. Lemma~\ref{lemma:scaling_property_3_LF}
summarizes the result.

\smallskip
\begin{lemma}\label{lemma:scaling_property_3_LF}
With a non-zero scale factor $|c|$, the critical delay $C_{\Pi}(n)$ under L\'{e}vy flight with
a distribution parameter $\alpha\in(0,2]$ scales as $\Omega(n^{\frac{\alpha}{2}})$.
\end{lemma}
\smallskip

Before proving the lemma, we give a remark. The scaling property of the critical delay with
continuous limit (shown in Theorem~\ref{thm:critical_delay_LF}) works as an upper bound for the
one without continuous limit. Hence, the result in Lemma~\ref{lemma:scaling_property_3_LF} shows
that our analysis in Section~\ref{sec:analysis:LF} gives the tightest upper bound, which justifies
our technique using continuous limit. We now give the proof of
Lemma~\ref{lemma:scaling_property_3_LF}.

\smallskip
\noindent\textit{Proof:} Similarly to the proof of Lemma~\ref{lemma:scaling_property_2_LF}, we will
prove this lemma by showing that
\begin{align}\label{eq:appendix:goal}
\lim_{n \to \infty} \pr\{T_x(c_d\sqrt{n}/\sqrt{2}) \le \tilde{t}(n)\} = 0,
\end{align}
where $\tilde{t}(n) = \Theta(n^{\frac{\alpha}{2}-\epsilon})$ for an arbitrary $\epsilon >0$ and
$\alpha\in(0,2]$. Then, from~(\ref{eq:lower_and_upper_prob}), we obtain
$\lim_{n \to \infty} \pr\{T(c_d\sqrt{n}) \le \tilde{t}(n)\} \le 2\lim_{n \to \infty}
\pr\{T_x(c_d\sqrt{n}/\sqrt{2}) \le \tilde{t}(n)\} = 0$. That is, we have $\lim_{n \to \infty}
\pr\{T(c_d\sqrt{n}) \le \tilde{t}(n)\} = 0$, or equivalently, $\lim_{n \to \infty}
\pr\{T(c_d\sqrt{n}) > \tilde{t}(n)\} = 1$, which shows that the critical delay $C_{\Pi}(n)$
scales as $\Omega(n^{\frac{\alpha}{2}})$.

Without loss of generality, we assume $\mathbf{X}(0) = (0,0)$. Then, from (\ref{eq:LF:location}),
$X_{x}(t)$ for $t=1,2,\ldots$ can be expressed as
\begin{align}\label{eq:appendix:location}
X_x(t) = \sum_{i=1}^{t} Z_i \cos \theta_i.
\end{align}
Let $Z\cos\theta$ denote the generic random variable for $Z_i\cos\theta_i$. By the independence
of random variables $Z$ and $\theta$, the mean of $Z\cos\theta$ is given by $\E [Z\cos\theta] =
\E [Z] \E [\cos\theta] = 0$. Hence, from (\ref{eq:appendix:location}), the mean of $X_x(t)$ is
given by $\E [X_x(t)] = t\E [Z\cos\theta]= 0$ and the variance of $X_x(t)$ becomes
$\E[(X_x(t))^2] = t\E[(Z\cos\theta)^2]$. Thus, Hoeffding's inequality~\cite{SL_Sharma07} gives an
upper bound for $\pr\{X_x(t) \ge r/\sqrt{2}\}$ as follows:
$
    \pr\{X_x(t) \ge r/\sqrt{2}\}
    \le \exp\left(-\frac{r^2}{8 t \E[(Z\cos\theta)^2]}\right).
$
By the symmetry of node motion, we have
\begin{align}\label{eq:appendix:bound_abs}
\pr\{|X_x(t)| \ge r/\sqrt{2}\} & \le 2 \exp\left(-\frac{r^2}{8 t \E[(Z\cos\theta)^2]}\right).
\end{align}
Due to (A1), the event $\{T_x(r/\sqrt{2}) \le k\}$ for $k=1,2,\ldots$ implies the event
$\bigcup_{t=1}^{k} \{|X_x(t)| \ge r/\sqrt{2}\}$. Hence, we have
\begin{align}\label{eq:appendix:HI_1}
\pr\{T_x(r/\sqrt{2}) \le k\}
& \le \sum_{t=1}^{k} \pr \{|X_x(t)| \ge r/\sqrt{2}\} \nonumber \\
& \le 2\sum_{t=1}^{k} \exp\left(-\frac{r^2}{8 t \E[(Z\cos\theta)^2]}\right)  \nonumber \\
& \le 2 k \exp\left(-\frac{r^2}{8 k \E[(Z\cos\theta)^2]}\right),
\end{align}
where the second inequality comes from (\ref{eq:appendix:bound_abs}). Substituting $r =
c_d\sqrt{n}$ and $k = \tilde{t}(n)$ into (\ref{eq:appendix:HI_1}), we have
\begin{align}\label{eq:appendix:HI_2}
\begin{split}
\lefteqn{\pr\{T_x(c_d\sqrt{n}/\sqrt{2}) \le \tilde{t}(n)\}} \\
& \quad \le 2 \tilde{t}(n) \exp\left(-\frac{(c_d)^2 n}{8 \tilde{t}(n)\E[(Z\cos\theta)^2]}\right).
\end{split}
\end{align}
In the following, we will derive a bound for $\E[(Z\cos\theta)^2]$. Since $\E[(Z\cos\theta)^2] =
\E[(Z|\cos\theta|)^2]$, we have
\begin{align}\label{eq:appendix:variance}
\E[(Z\cos\theta)^2]
&= \int_{0}^{\sqrt{n}} z^2 \text{d}F_{Z|\cos\theta|}(z).
\end{align}
We first consider the case of $\alpha\in(0,2)$. From the CCDF of $Z|\cos\theta|$ given for $z\ge
1$ in (\ref{eq:LF:ccdf_2}), we have for $z\ge 1$,
\begin{align*}
\frac{\text{d}F_{Z|\cos\theta|}(z)}{\text{d}z} &= -\frac{\text{d}}{\text{d}z} \pr\{Z|\cos\theta|
> z\}\\
&= \frac{2 \alpha c(n)}{\pi z^{\alpha+1}} \int_{0}^{\cos^{-1}(\frac{z}{\sqrt{n}})}
(\cos\vartheta)^\alpha \text{d}\vartheta \\
&\le \frac{\alpha c^\star c(n)}{z^{\alpha+1}}.
\end{align*}
Thus, the integral on the right-hand side of (\ref{eq:appendix:variance}) is bounded above by
\begin{align*}
\lefteqn{\int_{0}^{\sqrt{n}} z^2 \text{d}F_{Z|\cos\theta|}(z)}\\
&= \int_{0}^{1} z^2 \text{d}F_{Z|\cos\theta|}(z) + \int_{1}^{\sqrt{n}}  z^2
\text{d}F_{Z|\cos\theta|}(z) \\
&\le \pr\{0 \le Z|\cos\theta| \le 1\} + \int_{1}^{\sqrt{n}} z^2 \frac{\alpha c^\star
c(n)}{z^{\alpha+1}} \text{d}z \\
&= \pr\{0 \le Z|\cos\theta| \le 1\} + \frac{\alpha c^\star c(n)}{2-\alpha}
(n^{1-\frac{\alpha}{2}}-1),
\end{align*}
from which we have
\begin{align}\label{eq:appendix:order1}
\E[(Z\cos\theta)^2]= O(n^{1-\frac{\alpha}{2}}) \quad \text{ for } \alpha\in(0,2).
\end{align}
We next consider the case of $\alpha = 2$. By following the approach in the case of
$\alpha\in(0,2)$, we have
\begin{align}\label{eq:appendix:order2}
\E[(Z\cos\theta)^2]= O(n^{1-\frac{\alpha}{2}}) \quad \text{ for } \alpha=2.
\end{align}
Combining (\ref{eq:appendix:order1}) and (\ref{eq:appendix:order2}) gives $\E[(Z\cos\theta)^2]= O(n^{1-\frac{\alpha}{2}})$ for $\alpha\in(0,2]$. Hence, there exist
constants $\bar{n}\in\mathbb{N}$ and $\bar{c} > 0$ such that
\begin{align}\label{eq:appendix:bd1}
\E[(Z\cos\theta)^2] \le \bar{c} n^{1-\frac{\alpha}{2}} \quad \text { for all  } n \ge \bar{n}.
\end{align}
In addition, since $\tilde{t}(n)=\Theta(n^{\frac{\alpha}{2} - \epsilon})$, there exist constants
$\tilde{n}\in\mathbb{N}$ and $\tilde{c}>0$ such that
\begin{align}\label{eq:appendix:bd2}
\tilde{t}(n) \le \tilde{c} n^{\frac{\alpha}{2}-\epsilon} \quad \text{ for all } n\ge \tilde{n}.
\end{align}
By (\ref{eq:appendix:bd1}) and (\ref{eq:appendix:bd2}), the term on the right-hand side of
(\ref{eq:appendix:HI_2}) is further bounded by
\begin{align}\label{eq:appendix:bd3}
\begin{split}
\lefteqn{2 \tilde{t}(n) \exp\left(-\frac{(c_d)^2 n}{8 \tilde{t}(n)\E[(Z\cos\theta)^2]}\right)} \\
& \quad \le 2 \tilde{c} n^{\frac{\alpha}{2}-\epsilon} \exp\left(-\frac{(c_d)^2
n^{\epsilon}}{8\tilde{c}\bar{c}} \right).
\end{split}
\end{align}
By L'H\^{o}spital's rule, (\ref{eq:appendix:bd3}) becomes in the limit as
\begin{align}\label{eq:appendix:bd4}
\begin{split}
\lefteqn{\lim_{n\to\infty} 2 \tilde{t}(n) \exp\left(-\frac{(c_d)^2 n}{8
\tilde{t}(n)\E[(Z\cos\theta)^2]}\right)}\\
& \quad \le \lim_{n\to\infty} 2 \tilde{c}
n^{\frac{\alpha}{2}-\epsilon} \exp\left(-\frac{(c_d)^2  n^{\epsilon}}{8\tilde{c}\bar{c}} \right)
=0.
\end{split}
\end{align}
Combining (\ref{eq:appendix:HI_2}) and (\ref{eq:appendix:bd4}) proves (\ref{eq:appendix:goal}).
This completes the proof. \hfill $\square$

\section*{Acknowledgements}
This work is in part supported by the KRCF (Korea Research Council of Fundamental Science and Technology and the MKE(The Ministry of Knowledge Economy), Korea, under the ITRC(Information Technology Research Center) support program supervised by the NIPA(National IT Industry Promotion Agency) (NIPA-2010-(C1090-1011-0004)). This work is also in part supported by National Science Foundation under grants CNS-0910868, CNS-1016216 and the U.S. Army Research Office (ARO) under grant W911NF-08-1-0105 managed by NCSU Secure Open Systems Initiative (SOSI).

\begin{IEEEbiography}
[{\includegraphics[width=1in,height=1.25in,keepaspectratio]{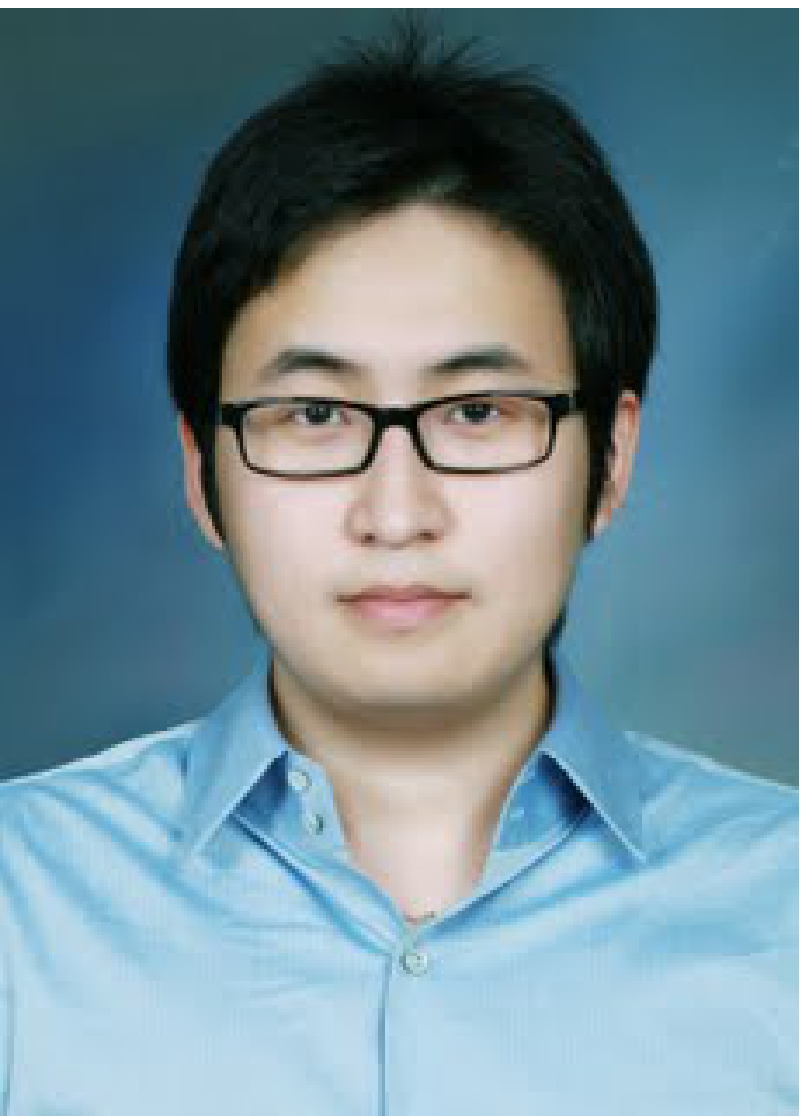}}]
{Kyunghan Lee} (S'07-M'10) received his B.S., M.S. and Ph.D. degrees from Electrical Engineering and Computer Science from Korea Advanced Institute of Science and Technology (KAIST), Daejeon, Korea in 2002, 2004 and 2009, respectively. He is currently a Senior Research Scholar in the Department of Computer Science at North Carolina State University. His research interests are in the areas of human mobility, delay tolerant networks, context-aware services and applications, mobile system and protocol design.
\end{IEEEbiography}

\begin{IEEEbiography} [{\includegraphics[width=1in,height=1.25in,keepaspectratio]{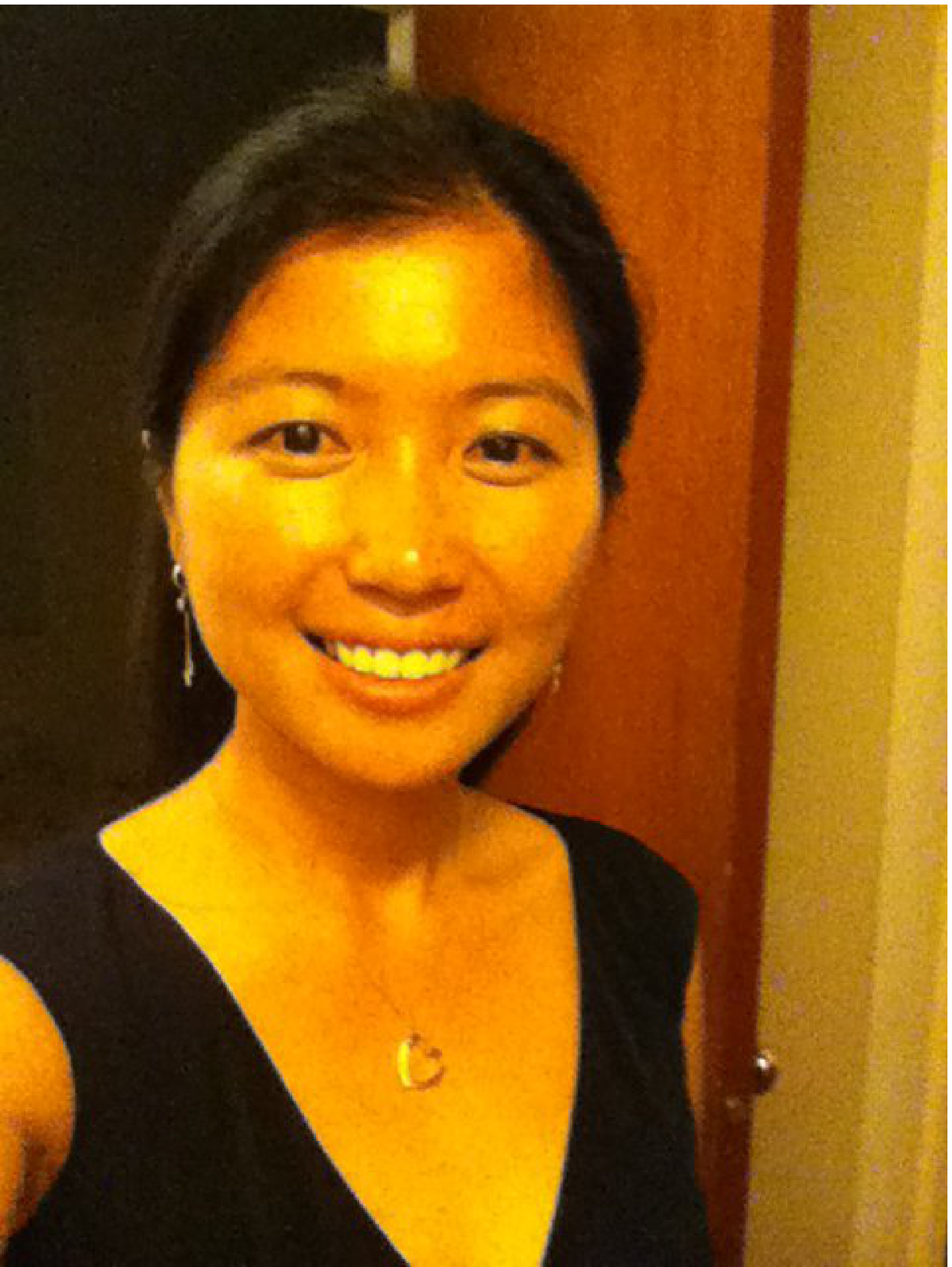}}]
{Yoora Kim} (S'05-M'09) received her B.S., M.S. and Ph.D. degrees in Mechanical Engineering, Applied Mathematics and Mathematical Sciences from Korea Advanced Institute of Science and Technology (KAIST), Daejeon, Korea, in 2003, 2005 and 2009, respectively. She is currently a post-doctoral research associate at The Ohio State University. Her research interests include modeling, design, and performance evaluation of communication systems, and scheduling and resource allocation problems in wireless networks under various stochastic dynamics.
\end{IEEEbiography}

\begin{IEEEbiography}
[{\includegraphics[width=1in,height=1.25in,keepaspectratio]{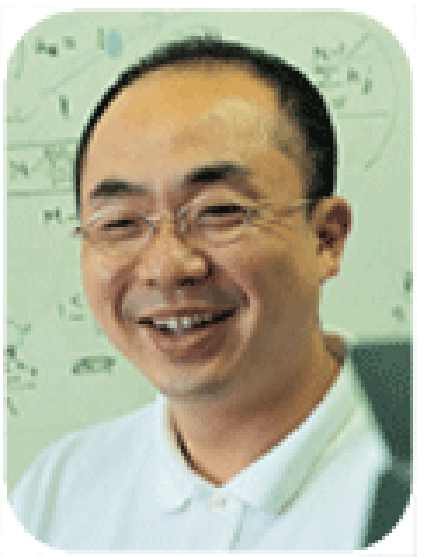}}]
{Song Chong} (M'93) received the B.S. and M.S. degrees in Control and Instrumentation Engineering from Seoul National University, Seoul, Korea, in 1988 and 1990, respectively, and the Ph.D. degree in Electrical and Computer Engineering from the University of Texas at Austin in 1995. Since March 2000, he has been with the Department of Electrical Engineering, Korea Advanced Institute of Science and Technology (KAIST), Daejeon, Korea, where he is a Professor and was the Head of the Communications and Computing Group of the department.

Prior to joining KAIST, he was with the Performance Analysis Department, AT\&T Bell Laboratories, New Jersey, as a Member of Technical Staff. His current research interests include wireless networks, future Internet, and human mobility characterization and its applications to mobile networking. He has published more than 100 papers in international journals and conferences.

He is an Editor of Computer Communications journal and Journal of Communications and Networks. He has served on the Technical Program Committee of a number of leading international conferences including IEEE INFOCOM and ACM CoNEXT. He serves on the Steering Committee of WiOpt and was the General Chair of WiOpt'09. He is currently the Chair of Wireless Working Group of the Future Internet Forum of Korea and the Vice President of the Information and Communication Society of Korea.
\end{IEEEbiography}

\begin{IEEEbiography}
[{\includegraphics[width=1in,height=1.25in,keepaspectratio]{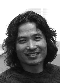}}]
{Injong Rhee} (S'89-M'94) received his Ph.D. from the University of North Carolina at Chapel Hill. He is a Professor in the Department of Computer Science at North Carolina State University. He is an Editor of IEEE Transactions on Mobile Computing. His areas of research interests include computer networks, congestion control, wireless ad hoc networks and sensor networks.
\end{IEEEbiography}

\begin{IEEEbiography}
[{\includegraphics[width=1in,height=1.25in,keepaspectratio]{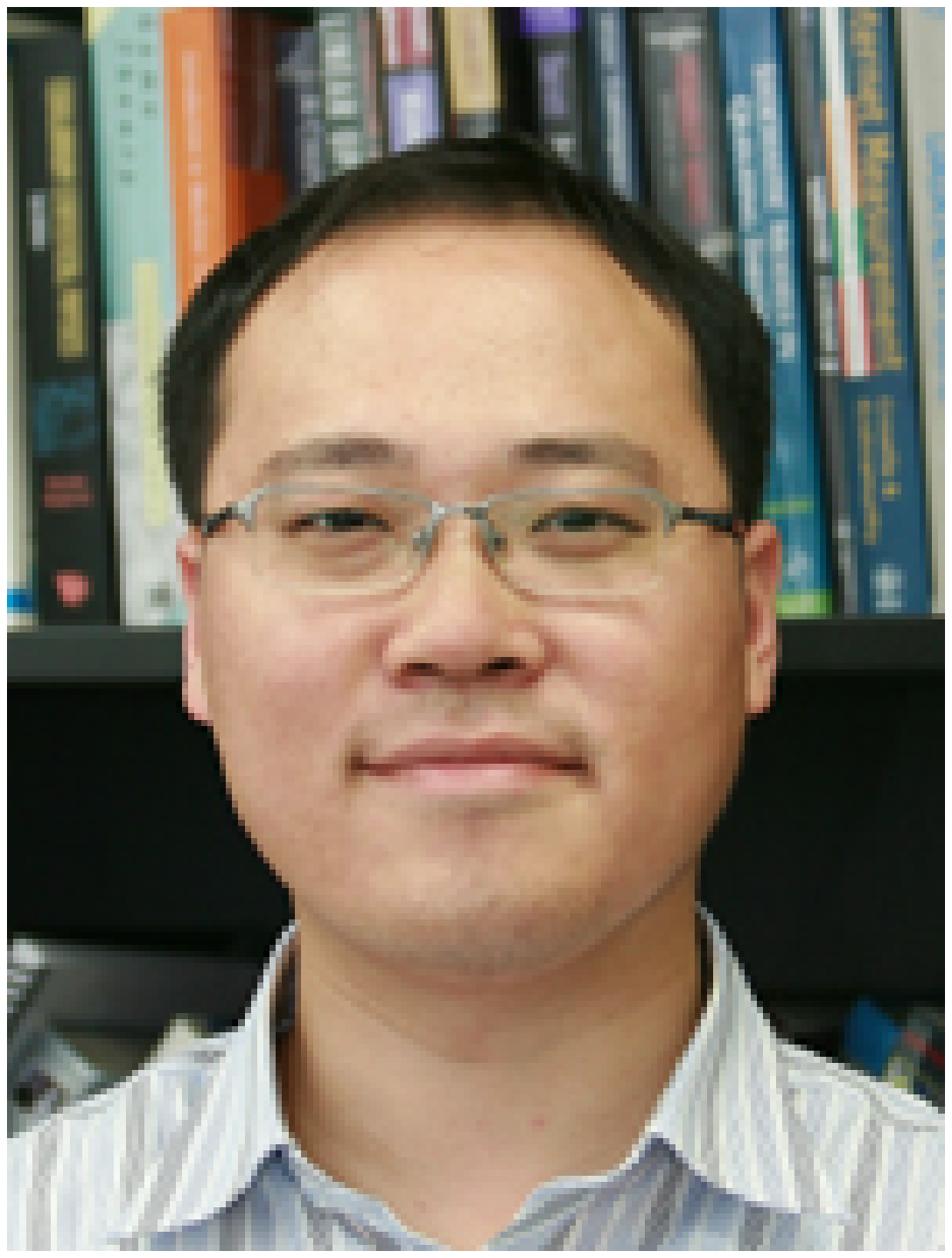}}]
{Yung Yi} (S'04-M'06) received his B.S. and the M.S. in the School of Computer Science and Engineering from Seoul National University, South Korea in 1997 and 1999, respectively, and his Ph.D. in the Department of Electrical and Computer Engineering at the University of Texas at Austin in 2006. From 2006 to 2008, he was a post-doctoral research associate in the Department of Electrical Engineering at Princeton University. Now, he is an associate professor at the Department of Electrical Engineering at KAIST, South Korea. He has been serving as a TPC member at various conferences including ACM Mobihoc, Wicon, WiOpt, IEEE Infocom, ICC, Globecom, and ITC. His academic service also includes the local arrangement chair of WiOpt 2009 and CFI 2010, the networking area track chair of TENCON 2010, and the publication chair of CFI 2010, and a guest editor of the special issue on Green Networking and Communication Systems of IEEE Surveys and Tutorials. He also serves as the co-chair of the Green Multimedia Communication Interest Group of the IEEE Multimedia Communication Technical Committee. His current research interests include the design and analysis of computer networking and wireless systems, especially congestion control, scheduling, and interference management, with applications in wireless ad hoc networks, broadband access networks, economic aspects of communication networks economics, and greening of network systems.
\end{IEEEbiography}

\begin{IEEEbiography}
[{\includegraphics[width=1in,height=1.25in,keepaspectratio]{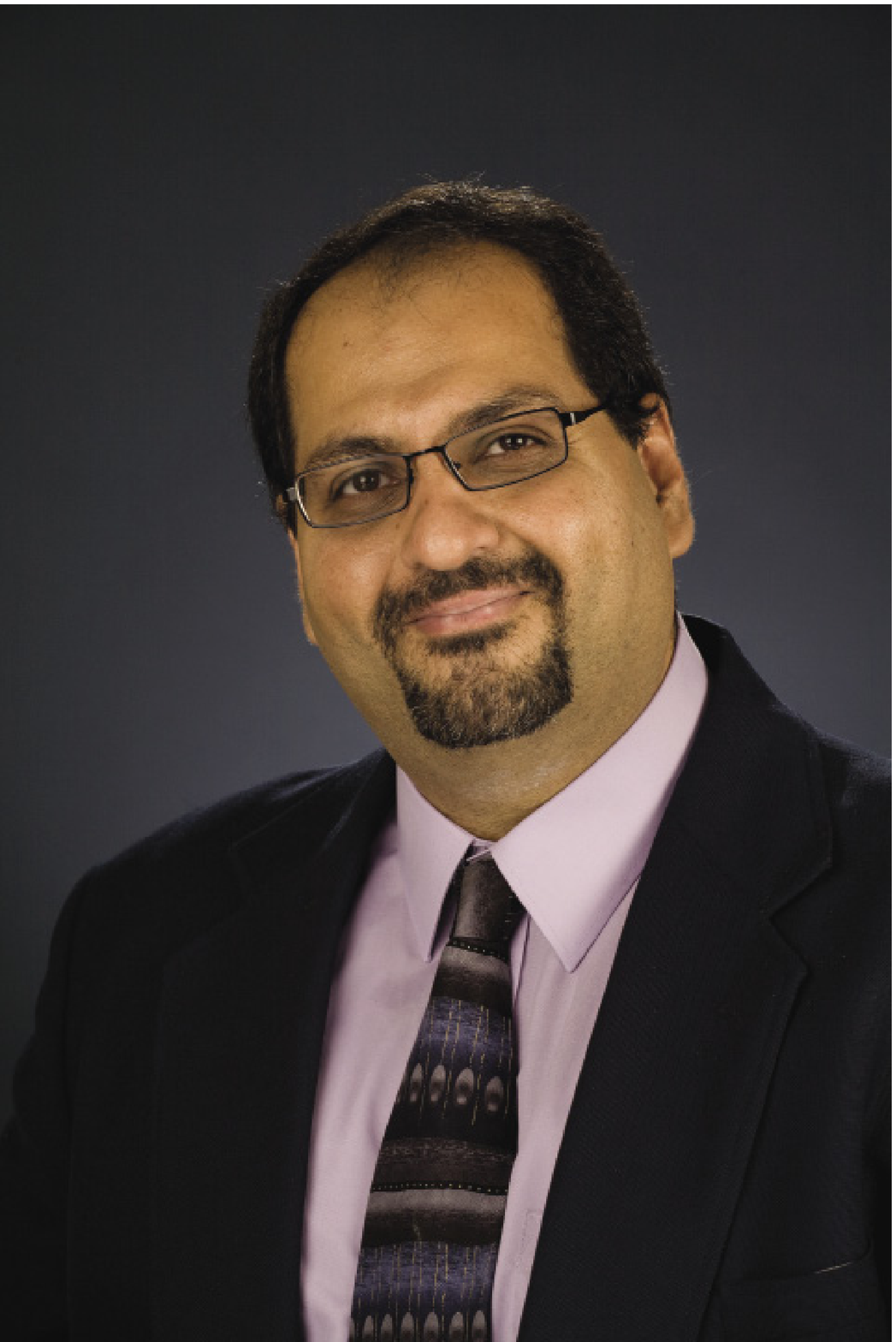}}]
{Ness B. Shroff} (S'91-M'93-SM'01-F'07) received the Ph.D. degree from Columbia University, New York, NY, in 1994. He currently holds the Ohio Eminent Scholar chaired professorship in Networking and Communications in the departments of ECE and CSE at The Ohio State University, in Columbus, OH. He also currently serves as a Guest Chaired Professor of wireless communications with the Department of Electronic Engineering, Tsighnua University, Beijing, China. Previously, he was a Professor of ECE with Purdue University, West Lafayette, IN, and the Director of the Center for Wireless Systems and Applications (CWSA), a university-wide center on wireless systems and applications. His research interests span the areas of communications, social, and cyber-physical networks, where he investigates fundamental problems in the design, performance, pricing, and security of these networks. Dr. Shroff has received numerous awards for his networking research, including the National Science Foundation CAREER award, the Best Paper awards for IEEE INFOCOM 2006 and 2008, the Best Paper Award for IEEE IWQoS 2006, the Best Paper of the Year Award for Computer Networks, and the Best Paper of the Year Award for the Journal of Communications and Networks (his IEEE INFOCOM 2005 paper was one of two runner-up papers).
\end{IEEEbiography}

\begin{thebibliography}{10}
\bibitem{SL1Gupta}
P.~Gupta and P.~R. Kumar, ``The capacity of wireless networks,'' \emph{IEEE
  Transaction on Information Theory}, vol.~46, no.~2, pp. 388--404, 2000.

\bibitem{grossglauser:mobility}
M.~Grossglauser and D.~N.~C. Tse, ``Mobility increases the capacity of ad hoc
  wireless networks,'' \emph{IEEE/ACM Transactions on Networking}, vol.~10,
  no.~4, pp. 477--486, 2002.

\bibitem{SL17Toumpis}
S.~Toumpis and A.~Goldsmith, ``Large wireless networks under fading, mobility,
  and delay constraints,'' in \emph{Proceedings of IEEE INFOCOM}, 2004.

\bibitem{SL14Neely}
M.~Neely and E.~Modiano, ``Capacity and delay tradeoffs for ad-hoc mobile
  networks,'' \emph{IEEE Transaction on Information Theory}, vol.~51, no.~6,
  pp. 1917--1937, 2005.

\bibitem{SL21Lin}
X.~Lin, G.~Sharma, R.~R. Mazumdar, and N.~B. Shroff, ``Degenerate
  delay-capacity tradeoffs in ad-hoc networks with brownian mobility,''
  \emph{IEEE/ACM Transactions on Networking}, vol.~14, pp. 2777--2784, 2006.

\bibitem{SL_Sharma07}
G.~Sharma, R.~Mazumdar, and N.~B. Shroff, ``Delay and capacity trade-offs in
  mobile ad hoc networks: a global perspective,'' \emph{IEEE/ACM Transactions
  on Networking}, vol.~15, no.~5, pp. 981--992, 2007.

\bibitem{rhee:levymobility}
I.~Rhee, M.~Shin, S.~Hong, K.~Lee, and S.~Chong, ``On the levy walk nature of
  human mobility,'' in \emph{Proceedings of INFOCOM}, 2008.

\bibitem{lee:slaw09}
K.~Lee, S.~Hong, S.~Kim, I.~Rhee, and S.~Chong, ``Slaw: A new human mobility
  model,'' in \emph{Proceedings of IEEE INFOCOM}, 2009.

\bibitem{gonalez:understanding}
M.~C. Gonzalez, C.~A. Hidalgo, and A.-L. Barabasi, ``Understanding individual
  human mobility patterns,'' \emph{Nature}, vol. 453, pp. 779--782, June 2008.

\bibitem{shlesinger:levyDynamics}
M.~F. Shlesinger, G.~M. Zaslavsky, and J.~Klafter, ``Levy dynamics of enhanced
  diffusion: Application to turbulence,'' \emph{Physical Review Letters},
  vol.~58, pp. 1100--1103, March 1987.

\bibitem{SL4Franceschetti}
M.~Franceschetti, O.~Dousse, D.~Tse, and P.~Thiran, ``Closing the gap in the
  capacity of wireless networks via percolation theory,'' \emph{IEEE
  Transactions on Information Theory}, vol.~53, no.~3, pp. 1009--1018, 2007.

\bibitem{SL5Ashish}
A.~Agarwal and P.~R. Kumar, ``Capacity bounds for ad hoc and hybrid wireless
  networks,'' \emph{ACM SIGCOMM Computer Communication Review}, vol.~34, no.~3,
  pp. 71--81, 2004.

\bibitem{SL8Bansal}
N.~Bansal and Z.~Liu, ``Capacity, delay and mobility in wireless ad-hoc
  networks,'' in \emph{Proceedings of IEEE INFOCOM}, 2003.

\bibitem{SL12Perevalov}
E.~Perevalov and R.~Blum, ``Delay limited capacity of ad hoc networks:
  Asymptotically optimal transmission and relaying strategy,'' in
  \emph{Proceedings of IEEE INFOCOM}, 2003.

\bibitem{SL13Tsingos}
A.~Tsirigos and Z.~J. Haas, ``Multipath routing in the presence of frequent
  topological changes,'' \emph{IEEE Communication Magazine}, vol.~39, no.~11,
  pp. 132--138, 2001.

\bibitem{SL18Sharma}
G.~Sharma and R.~Mazumdar, ``Scaling laws for capacity and delay in wireless ad
  hoc networks with random mobility,'' in \emph{Proceedings of IEEE ICC}, 2004.

\bibitem{SL20Lin}
X.~Lin and N.~B. Shroff, ``The fundamental capacity-delay tradeoff in large
  mobile ad hoc networks,'' in \emph{Proceedings of 3rd Annual Mediterranean Ad
  Hoc Networking Workshop}, 2004.

\bibitem{SL16Neely}
M.~Neely and E.~Modiano, ``Dynamic power allocation and routing for satellite
  and wireless networks with time varying channels,'' in \emph{Ph.D Thesis,
  Massachusetts Institute of Technology}, 2004.

\bibitem{SL22Gamal}
A.~El~Gamal, J.~Mammen, B.~Prabhakar, and D.~Shah, ``Optimal throughput-delay
  scaling in wireless networks: {Part I}: the fluid model,'' \emph{IEEE/ACM
  Transactions on Networking}, vol.~14, pp. 2568--2592, 2006.

\bibitem{RW01Drysdale}
P.~M. Drysdale and P.~A. Robinson, ``L\'{e}vy random walks in finite systems,''
  \emph{Physical Review E}, vol.~58, no.~5, pp. 5382--5394, 1998.

\bibitem{RW05Metzler}
R.~Metzler and J.~Klafter, ``The random walk's guide to anomalous diffusion: A
  fractional dynamics approach,'' \emph{Physics Reports}, vol. 339, no.~1, pp.
  1--77, 2000.

\bibitem{Chechkin2006}
A.~V. Chechkin, V.~Y. Gonchar, J.~Klafter, and R.~Metzler, ``Fundamentals of
  l\'{e}vy flight processes,'' \emph{Advances in Chemical Physics}, vol. 133,
  pp. 439--496, 2006.

\bibitem{nolan:2012}
J.~P. Nolan, \emph{Stable Distributions - Models for Heavy Tailed Data}.\hskip
  1em plus 0.5em minus 0.4em\relax Boston: Birkhauser, 2012, in progress,
  Chapter 1 online at academic2.american.edu/$\sim$jpnolan.

\bibitem{PhysRevE.73.057102}
M.~Ferraro and L.~Zaninetti, ``Mean number of visits to sites in levy
  flights,'' \emph{Physical Review E}, vol.~73, May 2006.

\bibitem{RW09Redner}
S.~Redner, \emph{A Guide to First-passage Processes}.\hskip 1em plus 0.5em
  minus 0.4em\relax New York: Cambridge University Press, 2001.

\bibitem{RW04Gitterman}
M.~Gitterman, ``Mean first passage time for anomalous diffusion,''
  \emph{Physical Review E}, vol.~62, no.~5, pp. 6065--6070, 2000.

\bibitem{Podlubn1999}
I.~Podlubny, \emph{Fractional Differential Equations}.\hskip 1em plus 0.5em
  minus 0.4em\relax London: Academic Press, 1999.

\bibitem{RW06Marsden}
J.~E. Marsden and M.~J. Hoffman, \emph{Elementary Classical Analysis}.\hskip
  1em plus 0.5em minus 0.4em\relax New York: W. H. Freeman and Company, 1993.

\bibitem{Ross1996}
S.~M. Ross, \emph{Stochastic Processes}.\hskip 1em plus 0.5em minus 0.4em\relax
  New York: John Wiley \& Sons, 1996.

\bibitem{SL33Garetto}
M.~Garetto, P.~Giaccone, and E.~Leonardi, ``Capacity scaling in ad hoc networks
  with heterogeneous mobile nodes: the super-critical regime,'' \emph{IEEE/ACM
  Transactions on Networking}, vol.~17, no.~5, pp. 1522--1535, 2009.

\bibitem{SL35Alfano_j}
G.~Alfano, M.~Garetto, E.~Leonardi, and V.~Martina, ``Capacity scaling of
  wireless networks with inhomogeneous node density: Lower bounds,''
  \emph{IEEE/ACM Transactions on Networking}, vol.~18, no.~5, pp. 1624--1636,
  2010.
\end{thebibliography}
\end{document}